\documentclass[letterpaper,conference]{IEEEtran}
\usepackage{graphicx,graphics}
\usepackage{comment}
\usepackage{epstopdf}
\usepackage[ruled,vlined,linesnumbered]{algorithm2e}
\usepackage{cite}
\usepackage{multirow}
\usepackage{soul,color}
\usepackage{ulem}

\ifCLASSINFOpdf
\DeclareGraphicsExtensions{.pdf,.eps}
\else
\fi

\usepackage[fleqn]{amsmath}
\usepackage[tight,footnotesize]{subfigure}

\usepackage{url}
\usepackage{amsfonts}
\usepackage{amsthm}
\usepackage{tabularx}
\usepackage[printonlyused]{acronym}
\usepackage{framed}
\usepackage[draft,footnote,nomargin]{fixme}

\newcolumntype{v}[1]{>{\raggedright\hspace{0pt}}p{#1}}
\renewcommand{\arraystretch}{1.2}

\newcommand{\fref}[1]{Fig.~\ref{#1}}
\newcommand{\sref}[1]{Section~\ref{#1}}
\newcommand{\tref}[1]{Table~\ref{#1}}
\newcommand{\eref}[1]{Eq.~\ref{#1}}

\begin{document}
	\normalem
	
\acrodef{CDF}{Cumulative Density Function}
\acrodef{EPC}{Evolved Packet Core}
\acrodef{mRRH}{\emph{master-} Remote Radio Head}
\acrodef{RRH}{Remote Radio Head}
\acrodef{mAN}{master Access Node}
\acrodef{AN}{Access Node}
\acrodef{MCS}{Modulation and Coding Scheme}
\acrodef{COMP}{Cooridnated Multi-Point}
\acrodef{QoS}{Quality of Service}
\acrodef{IoT}{Internet of Things}
\acrodef{AP}{Access Point}
\acrodef{AI}{Artificial Intelligence}
\acrodef{CBR}{Case Based Reasoning}
\acrodef{OBS}{Ontology Based System}
\acrodef{RBS}{Rule Based System}
\acrodef{GA}{Genetic Algorithm}
\acrodef{SA}{Simulated Annealing}
\acrodef{LS}{Local Search}
\acrodef{TS}{Tabu Search}
\acrodef{CCA}{Clear Channel Assessment}
\acrodef{RTT}{Round Trip Time}
\acrodef{RSSI}{Received Signal Strength Indicator}
\acrodef{KB}{Knowledge Base}
\acrodef{QoS}{Quality of Service}
	
	\bibliographystyle{IEEEtran}
	
	\IEEEoverridecommandlockouts	
	
	\title{Artificial Intelligence Inspired Self-Deployment of Wireless Networks}
	

\author{
	\IEEEauthorblockN{Erma Perenda\IEEEauthorrefmark{1}, Ramy Atawia\IEEEauthorrefmark{2} and Haris Gacanin\IEEEauthorrefmark{2}}
	\IEEEauthorblockA{\IEEEauthorrefmark{1}Nokia Shanghai Bell, China}
		\IEEEauthorblockA{\IEEEauthorrefmark{2}Nokia Bell Labs, Belgium, haris.gacanin@nokia-bell-labs.com}
		\\ [-4.0ex]
	}
	
\maketitle
\begin{abstract}
In this paper, we propose a self-deployment approach for finding the optimal placement of extenders in which both the wireless back-haul and front-haul throughput of the extender are optimized. We present an artificial intelligence (AI) case based reasoning (CBR) framework that enables autonomous self-deployment in which the network can learn the environment by means of sensing and perception. New actions, i.e. extender positions, are created by problem-specific optimization and semi-supervised learning algorithms that balance exploration and exploitation of the search space. An IEEE 802.11 standard compliant simulations are performed to evaluate the framework on a large scale and compare its performance against existing conventional coverage maximization approaches. Experimental evaluation is also performed in an enterprise environment to demonstrate the competence of the proposed AI-framework in perceiving such a dense scenario and reason the extender deployment that achieves user quality of service (QoS). Throughput fairness and ubiquitous QoS satisfaction are achieved which provide a leap to apply AI-driven self-deployment in wireless networks.
\end{abstract}

\begin{keywords}
Artificial intelligence, optimization, wireless network.
\end{keywords}

\section{{Introduction}}
	\label{sec:Intro}
IEEE 802.11 wireless network is expected to serve more than 50\% of the global data traffic in 2021 \cite{ciscoVNI}. Such a network will thus employ a large number of access points (APs) with wired backhaul\footnote{In this paper, wireless APs with wired backhaul is refered as master AP (mAP).}, more than half billion, that are deployed in dynamic manner. At the same time, connectivity nodes having a wireless back-haul, referred to as extenders (EXTs), are flooding the wireless indoor market to minimize the deployment cost and improve coverage \cite{ExtMarket}. Thus, shifting to multi-hop architectures, but increasing interference and contention \cite{bellalta2016ieee}. The main challenge in such deployments is the lack of coordination between mAP and EXT serving overlapping areas, and shared unlicenced spectrum by different network operators. New deployment strategy is crucial to achieve ubiquitous quality of service (QoS) satisfaction, and decrease the operational costs (e.g., number of help desk calls and on-site visits). The behaviour of uncoordinated neighbouring networks remains a game-changing factor, yet hard to be modelled by human rules. As such, autonomous self-deployment is attractive approach to ensure the optimality of positions in uncoordinated deployments at low operational costs.

This paper, for the first time in literature, introduces an artificial intelligence (AI) case based reasoning (CBR) framework for self-deployment of wireless network\footnote{The framework preliminarily appeared in our recent work in \cite{AIGC}.}. The framework enables network to sense the environment and build necessary knowledge used to 1) assess the optimality of current position and 2) propose new locations for EXTs. We design framework to ubiquitously monitor the user satisfaction and determine the optimality of current location. The network notifies the user to reposition an EXT to optimal location with guaranteed demand. The main contributions of this paper can be summarized as follows:
\begin{itemize}
\item We propose a general AI framework for self-deployment of wireless network based on CBR. The previous network states, optimization actions and rewards are frequently stored in the knowledge base (KB), and then used to guide the network while taking future decisions.
\item A problem-specific optimization with active learning is introduced to populate the KB with actions and their corresponding fitness values. This is done while tackling the search trade-off between exploitation and exploration. Hence, avoids trapping the search at local optimal solutions and prevent the network from revisiting discovered search space. Such methodology increases the chance of reaching optimal position of extenders at low searching and learning costs.
\item {To speed up learning and minimize the cost of learning we introduce semi-supervised learning to learn and adjust system variables (i.e. throughput) and exploration factor that controls optimization process -- more robust exploration and exploitation strategy. We introduce support vector machine (SVM) estimation of throughput variables when non-empty set of training data is present.}
\item We introduce the first testbed that integrates AI in wireless network and thus can be used as a baseline for self-deployment and other autonomous optimization techniques. In essence, the testbed adopts commercial off-the-shelf (COTS) devices with modified software that enabling integration with a remote management server hosting the framework in a distributed fashion enabling real-time monitoring. 
\item We adopt the IEEE 802.11ax standard compliant simulator ns-3 to evaluate the framework under typical home scenarios \cite{80211axSim}. In essence, the framework is evaluated on a larger scale using the simulator to address the corner cases and unveil performance bounds in a controlled environment which provides a benchmark for future self-deployment algorithms. The resultant QoS performance is compared against that of the existing techniques to demonstrate the competence of the proposed framework in perceiving the neighbouring environment and reasoning the optimal deployment.
\end{itemize}

\subsection{Related Works}
The concept of self-deployment previously appeared in cellular networks \cite{zhou2011fault}, \cite{SelfDeployment}, \cite{SelfDeploymentMSNObstacle} and \cite{SelfDeploymentVANET}. The optimal locations of base stations are recalculated and changed frequently according to the locations of hotspots and obstacles in order to satisfy capacity and coverage constraints. Wireless network self-deployment is more challenging due to shared (unlicensed) spectrum and non-uniform layouts creating coverage holes and hidden node problem.

Wireless deployment aims to find the minimal number and optimal locations of APs such that the network performance and user QoS are maximized either manually or computer-based \cite{ling2006joint}--\cite{IndoorPlanning_Heuristic}. The former refers to using a test hardware tool to perform a site survey for the indoor environment. Experienced network engineers temporarily deploy wireless transmitters (e.g. additional APs) in candidate positions to measure the coverage level. Based on human experience, different locations and number of transmitters are tested until the best coverage is achieved (e.g. no coverage holes). Although optimal coverage-based deployment can be achieved, the time taken during the survey further increases with the size of an environment (e.g. the number and complex layout of rooms) making the planning process prohibitively expensive.

As opposed to the survey-based approach, network operators adopted off-line planning software to calculate an optimal deployment. In essence, the software uses path loss models, traffic maps and building layout (e.g. number of rooms and type of walls) to find the optimal location and number of APs/EXTs by dividing the layout into square grids representing candidate locations for APs. Different configurations are tested by adding and removing the APs in the grids and detect their ability to satisfy the coverage, demand and other planning constraints. The software picks the configuration that satisfies all constraints and optimizes given network metric such as minimum cost or maximum throughput. Computer-based deployment is less expensive and more time-efficient than the survey-based approach. However the optimality is highly sensitive to the accuracy of path loss model being a function of different layouts - obtaining the layout for each scenario is unfeasible. Network operators tend to follow a conservative strategy by deploying more EXTs to compensate the errors in path loss model. Nevertheless, the traffic demand map might vary over time and the off-line calculated locations are no longer optimal.

 {The main challenges of existing approaches are a lack an algorithm to perceive the environment and high complexity of optimization problem. These challenges are tackled by using different heuristic optimization techniques (i.e., such as genetic algorithm (GA), simulated annealing (SA), local search (LS) and tabu search (TS)} \cite{ling2006joint,IndoorPlanning_Heuristic,zhou2011fault} {or developing guided heuristic} \cite{WLAN_Planning_Heuristic,WLAN_Planning_Heuristic_2}).

\subsection{Motivation for AI}
Wireless network deployment is considered as a double-edged sword. A network operator or a user may deploy an EXT in a position that extends the coverage of mAP, but does not necessarily improve QoS at the end-user location. In addition, the EXT shares radio resources with other users connected directly to the mAP. This makes a suboptimal placement of extender more challenging as it increases the risk of degrading the total system throughput \cite{Ext1}. Both survey-based and computer-based wireless deployments have complementary features, yet they face the following challenges:
\begin{enumerate}
\item \textit{Dynamics in the indoor environment:} The indoor environment typically experiences spatio-temporal variations in both demand and coverage. The spatial demand of users can change over the time due to utilizing services with different bandwidth or latency requirements. The coverage of deployed nodes varies due to activities of users (e.g. mobility) and orientation of devices \cite{WLANDeploymentManual}.
\item \textit{Neighboring network knowledge:} A neighboring network refers to another wireless system (i.e. another mAP-EXT pair) that is utilizing the same spectrum, but managed by another operator. While neighboring network knowledge is captured by the manual deployment, their spatio-temporal dynamics are not. In addition to coverage and demand variations in neighboring networks, other network operators might deploy new APs or extenders or reposition existing ones within the same geographical region causing changes in interference and contention behaviors.
\end{enumerate}

\begin{figure}[h]
	\centering
	\subfigure[Coverage]
	{\includegraphics[scale=0.28]{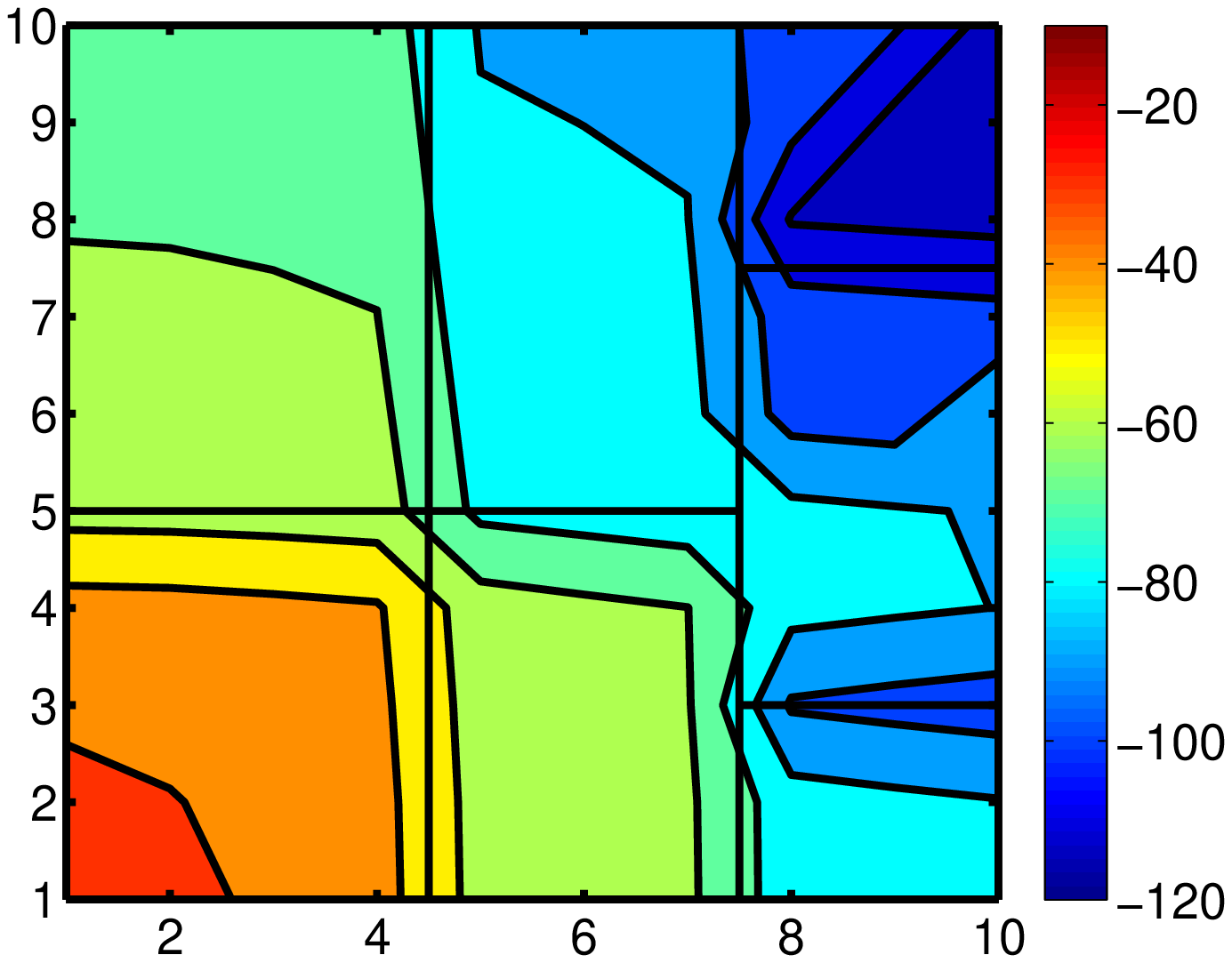}}\hspace{-0.8em}\vspace{-0.3em}%
	\label{fig:APRSSI}
	\subfigure[Throughput]
	{\includegraphics[scale=0.28]{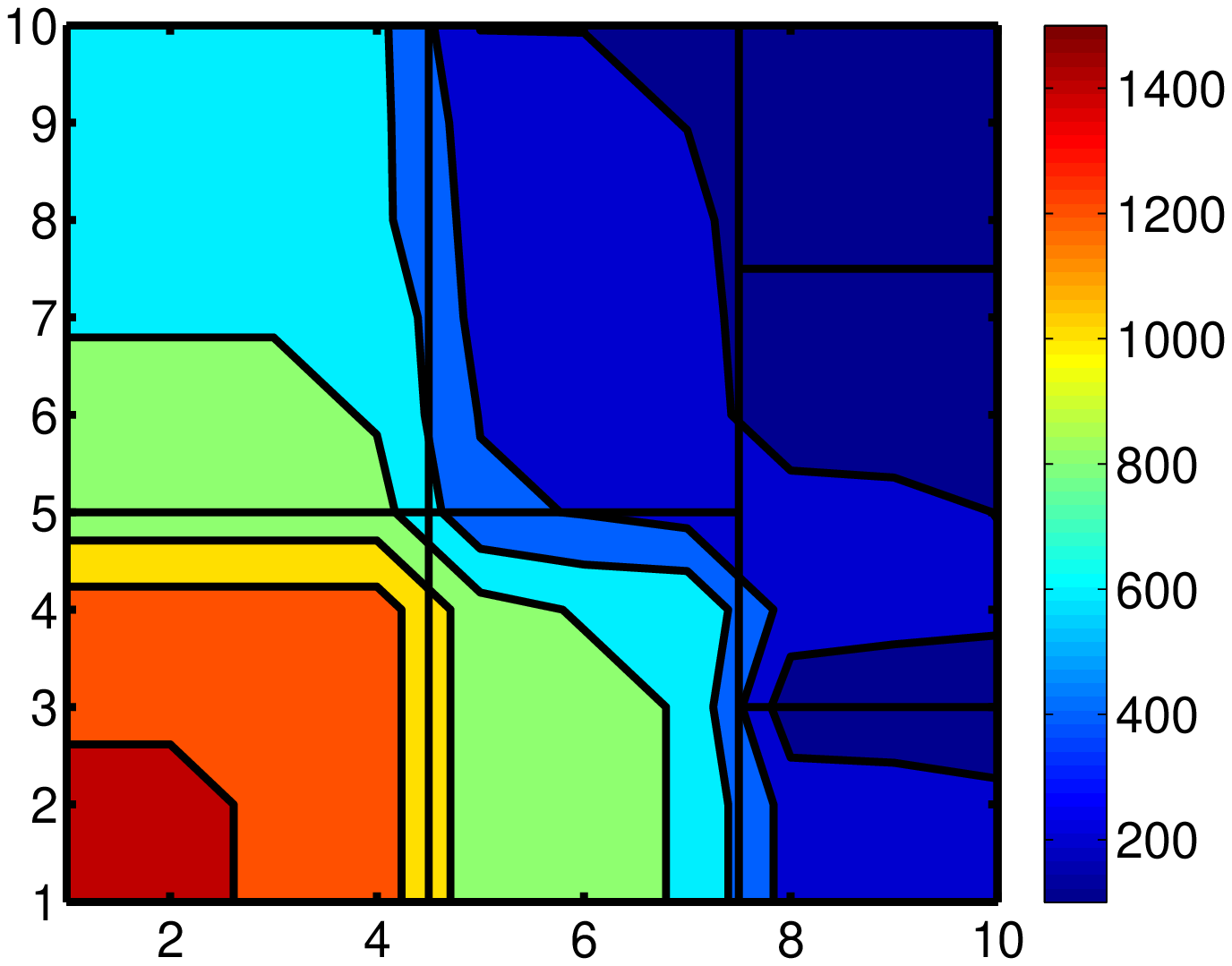}}\hspace{-1em}\vspace{-0.3em}%
	\label{fig:AP}
	
	\subfigure[Coverage]
	{\includegraphics[scale=0.28]{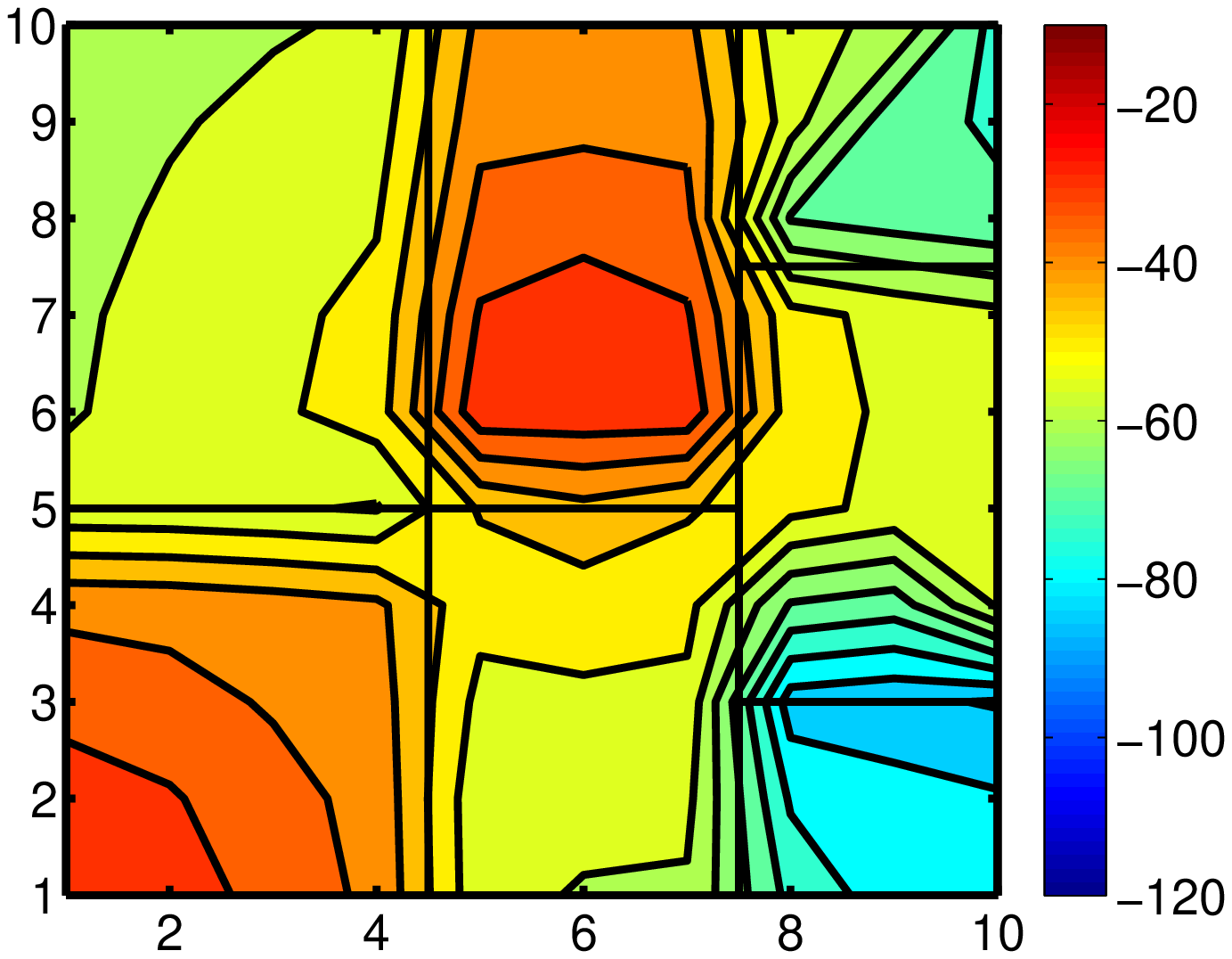}}\hspace{-0.8em}\vspace{-0.3em}%
	\label{fig:APExt1RSSI}
	\subfigure[Throughput]
	{\includegraphics[scale=0.28]{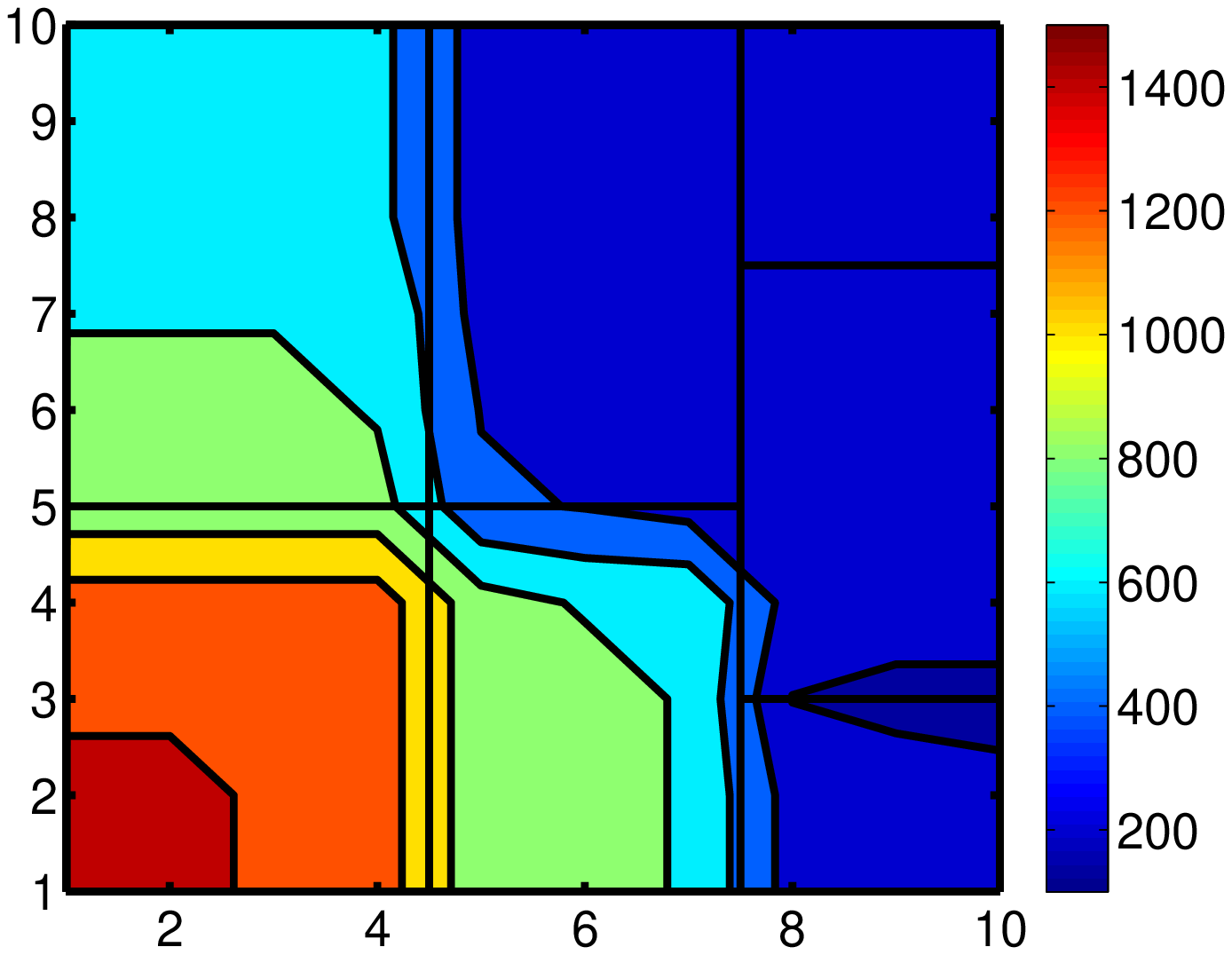}}\hspace{-1em}\vspace{-0.3em}%
	\label{fig:APExt1}

	\subfigure[Coverage]
	{\includegraphics[scale=0.28]{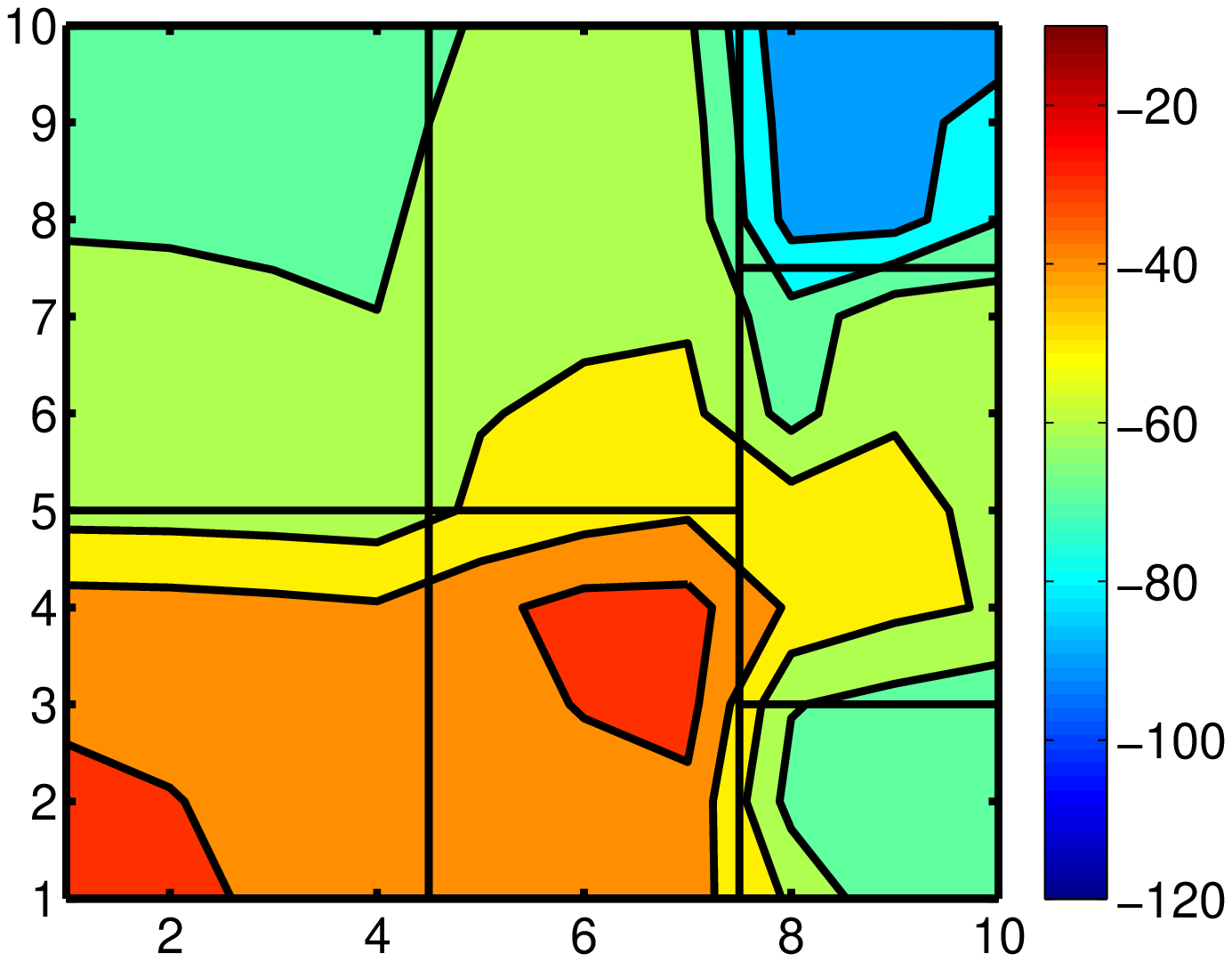}}\hspace{-0.8em}\vspace{-0.3em}%
	\label{fig:APExt3RSSI}
	\subfigure[Throughput]
	{\includegraphics[scale=0.28]{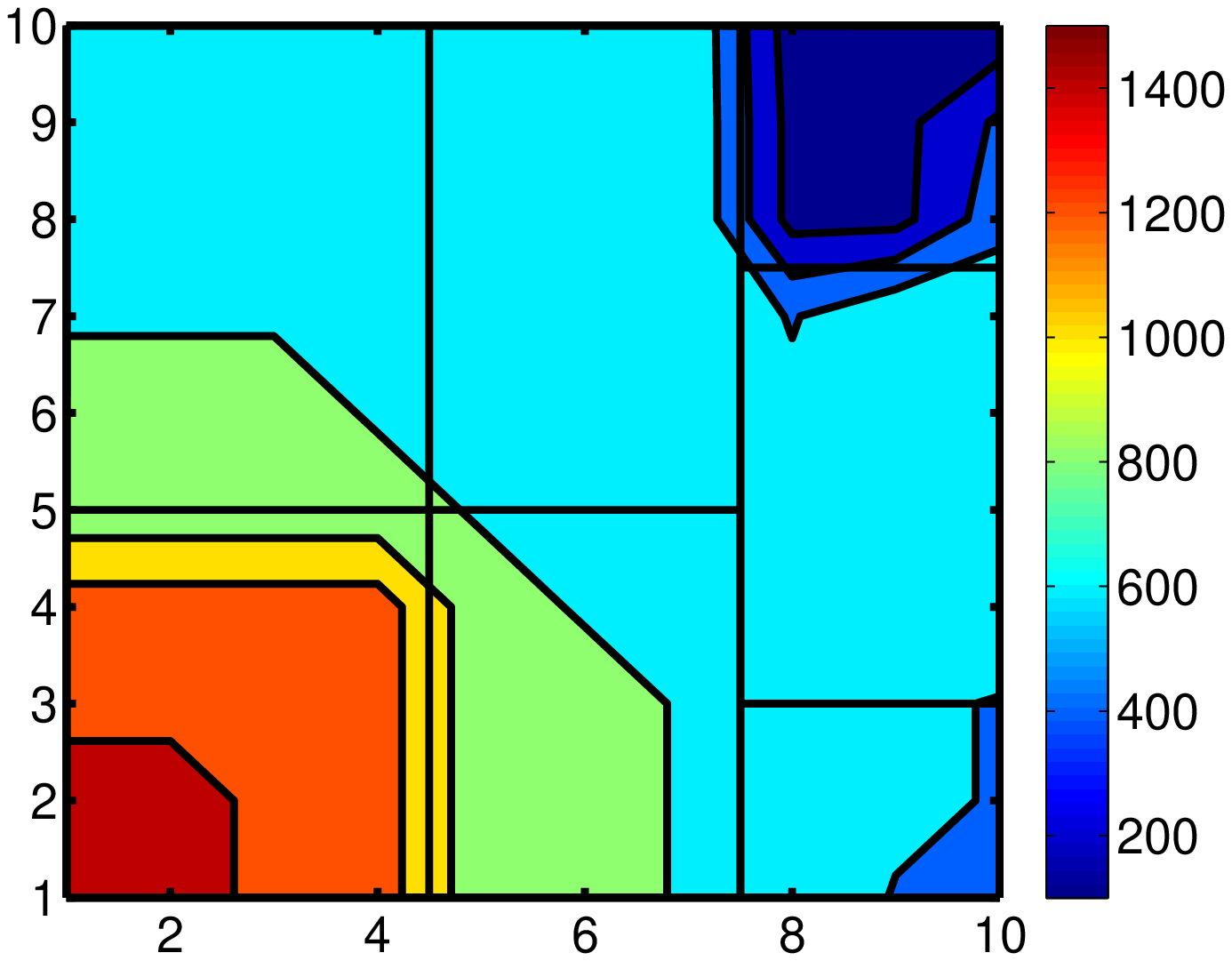}}\hspace{-1.0em}\vspace{-0.3em}%
	\label{fig:APExt3}
	\caption{Three indoor deployment scenarios where left and right figures, respectively, denote coverage and achievable user throughput in isolated apartment, with AP placed in position (1,1).}
	\label{fig:illust1}
\end{figure}

The problem at hand is to find the optimal locations of extenders associated to mAP such that the users' demands are satisfied. The optimal extender position has to balance the capacity on both links, the backhaul link (i.e. between mAP and EXT) and the fronthaul link (i.e. between EXT and user). In \fref{fig:illust1}, we present an isolated apartment scenario to illustrate the drawback of coverage based approach and the potential of AI-driven {self-deployment} proposed in this paper. Having mAP placed in the grid location (1,1) the following three scenarios are considered:
\begin{itemize}
\item \textit{Single AP per apartment} in \fref{fig:illust1}(a-b), where the position is restricted to the existing wired infrastructure, resulting in coverage holes illustrated in \fref{fig:illust1}(a) with low throughput values shown in \fref{fig:illust1}(b), which necessitates deployment of an extender;
\item \textit{Conventional coverage-driven deployment} in \fref{fig:illust1}(c-d) with the extender in the center of indoor area (i.e. midway between the mAP and coverage hole). Although the overall coverage is maximized as illustrated in \fref{fig:illust1}(c), such approach does not improve the throughput at user location as shown in \fref{fig:illust1}(d)). This is due to overlooking the back-haul link throughput, while only maximizing the front-haul link (extender-to-user). Even worse, inexperienced users will most likely place the extender closer to their devices, which reduces the back-haul throughput and consequently, limiting achievable service quality. Such solution is anticipated by the existing literature on EXT deployment that aim to minimize the packet transmission delay and maximize the coverage \cite{Ext1,Ext2}.
\item \textit{The proposed AI-driven self-deployment approach} is adopted in \fref{fig:illust1} (e-f) to sense and learn the environment, and reason the location that achieves the best compromise between back-haul and front-haul throughput. Hence, improves the total throughput as shown in \fref{fig:illust1}(f) while compromising the coverage as observed by comparing the \fref{fig:illust1}(c) and (e).
\end{itemize}
Hence, a novel self-deployment approach needs to capture the dynamics of neighbour's interference and contention, caused by variations of demand, placement of new APs or EXTs, that provokes the optimality of locations. 

\section{Network Model}\label{sec:netmod}
\subsection{System Model}
The system consists of single mAP and a group of extenders which are connected to mAP directly or through other extenders. The location of mAP is static and already determined, while locations of extenders can be changed through time controlled by the proposed AI framework. We assume that a new location is calculated by the framework and recommended to the user by a request at $ t \in\mathcal{T} $. 

{We note that the framework can be hosted either on mAP or in the cloud taking decisions for a whole system. The framework collects the sensing data (i.e network metrics) and based on the current network metrics and historical data derives real-time solutions for extender relocation. The detailed system model is described by the following system variables.}
	
\subsubsection{Deployment Decision Variables}	
The location of each extender is changed based on the network conditions, and we define the following variables
\begin{itemize}
	\item $ \delta_{i,t} $ equals to 1 if the extender has to be deployed at location $ i \in \mathcal{I} $ after request $ t \in\mathcal{T} $; and equals 0 otherwise
	\item 	$ \alpha_{i,t} $ equals to 1 if an extender has to be removed from or deployed at location $ i $ after request $ t $; and equals 0 otherwise. This variable is used to track the number of extender repositions.
\end{itemize}

\subsubsection{Association Variables}
Each extender can be treated as a station associated to other extender or mAP. Hence, we define an association variable $ x_{i,t,u} $ which is equal to 1 if the station (extender or user) at location $ u $ is connected to the extender at location $ i $ after request $ t $; and equals 0 otherwise.

\subsubsection{{Throughput Variables}}
{The extender comprises of wireless connection with mAP (called backhaul link) and toward users (called fronthaul link) thus the corresponding throughput variables are defined as follows:}
\begin{itemize}
\item {$\hat{r}_{i,t}^{(b)}$ denotes the estimated throughput at the back-haul of extender deployed at location $ i $ after request $ t $. This variable is obtained as modulation and coding scheme (MCS) index} \cite{mcsindex} {having minimum difference from the maximum achievable throughput $C=B\times \log{(1+SNR_{i,t})}$, where $B$ denotes the channel width and $SNR_{i,t}$ denotes the signal-to-noise ratio at location $i$ after request $t$ determined based on free-space pathloss model.}
\item {$\bar{r}_{i,t}^{(b)}$ denotes the actual throughput measured at the back-haul of extender deployed at location $i$ after request $t$. Its value depends on the actual selected MCS and the decisions of MAC protocol (e.g. resource allocation and clear channel assessment (CCA)) obtained by the station statistics at an access point (mAP or other extender) to which target extender is connected. For example, either by vendor specific interface or through TR-069 protocol (i.e. InternetGatewayDevice. LANDevice. \{i\}. WLANConfiguration. \{i\}. AssociatedDevice. \{i\}. TxBitRate)} \cite{tr181}.
\item {$\hat{r}_{i,t,u}^{(f)}$ denotes the estimated front-haul throughput of the user at location $u$ and connected to extender or mAP at location $i$ after request $t$. This value is generated in the same way as $\hat{r}_{i,t}^{(b)}$ from above. Based on $\hat{r}_{i,t,u}^{(f)}$, the distance based front-haul throughput is calculated as $\hat{r}_{i,t}^{(f)}=\sum_{u=1}^U x_{i,t,u} \hat{r}_{i,t,u}^{(f)}$.}
\item {$\bar{r}_{i,t,u}^{(f)}$ denotes the actual throughput measured by the user at location $u$ and connected to extender located at $i$ after request $t$. This value depends on the actual selected MCS, and the MAC decisions (e.g. resource allocation and CCA) and it is calculated in the same way as $\bar{r}_{i,t}^{(b)}$ from the above. Based on $\bar{r}_{i,t,u}^{(f)}$, the distance based front-haul throughput is calculated as $\bar{r}_{i,t}^{(f)}=\sum_{u=1}^U x_{i,t,u} \bar{r}_{i,t,u}^{(f)}$, where $u$ is the user index.}
\item  {$r_{i,t}$ denotes E2E user throughput for all users connected to extender located at $i$ after request $t$. This value can be measured as $r_{i,t}=\min{(\bar{r}_{i,t}^{(b)},\bar{r}_{i,t}^{(f)})}$ or calculated as $r_{i,t}=\sum_{u=1}^U x_{i,t,u} r_{i,t,u}$, where $r_{i,t,u}=(TXBytes + RXBytes)\times 8/\Delta t$, where $TXBytes$ and $RXBytes$, respectively, denote the total number of bytes transmitted and received for the user $u$ within time interval $\Delta t$. For examples these values are available through specific vendor extensions (e.g. statistics counters InternetGatewayDevice. LANDevice.\{i\}.    WLANConfiguration.\{i\}. AssociatedDevice. \{i\}. Stats.BytesSent and InternetGatewayDevice.LANDevice. \{i\}. WLANConfiguration.\{i\}. AssociatedDevice.\{i\}. Stats.Bytes Received, respectively). 
Although the second way to obtain  E2E user throughput is more accurate it's drawback is requirement that the user devices are always active with the transmitting and receiving data requests.}

\end{itemize}

\subsubsection{{Demand Variables}}
{The demand of every user at location $ u $ after request $ t $ is denoted by $ D_{t,u} $ and represents the minimum throughput needed by the user to satisfy target QoS. The demand of each user can be obtained during the initial deployment of wireless system in scope of Service Level Agreement (SLA) between the user and a service provider. More advanced approach to obtain these variables would be by using traffic prediction techniques as in} \cite{joshi}, {but this is out of the scope of this paper. We assume that these variables are given by SLA.}

\subsection{Problem Formulation}
{As it is already said, the proposed AI framework is a centralized architecture which is aiming to find the optimal location of each extender in such way that each user demand is satisfied at each time instant. Based on that the problem is finding the optimal location(s) of extender with minimal cost.}  Unlike existing approaches, the framework does not have any prior knowledge about the network. As such, the layout of the building or wall losses are not adopted. This is in addition to the unavailability of neighbour information such as locations, channel configurations and traffic load. As such, the network will notify the user to change the location of extender to 1) learn the environment, and 2) improve his QoS level. It is therefore of paramount importance to minimize the number of requests to the user, i.e. ask the user to move the extender, which necessitates conscious and fast learning by the network. {The problem can be mathematically formulated as follows:}

\begin{equation}
\label{eq:CentralFormula1}
\underset{\mathbf{\alpha,\delta}}{\text{minimize}} \quad\,\,\, 
\left\{
\underset{\forall t \in T}{\text{max}} \sum_{\forall i\in I}
\delta_{i,t}+\sum_{\forall t\in T}\sum_{\forall i\in I}\alpha_{i,t}
\right\}
\end{equation}
\begin{equation*}
\begin{aligned}
\text{subject to:}&\\
\text{C1:}\quad  \alpha_{i,t} \geq |\delta_{i,t-1}-\delta_{i,t}|  \forall t \in\mathcal{T},\forall i \in\mathcal{I}& &\\
\text{C2:}\quad \sum_{i=0}^{I} \delta_{i,t} \bar{r}_{i,t,u}^{(f)} \geq {D}_{t,u},  \forall t \in\mathcal{T},\forall u \in\mathcal{U},& &\\
\text{C3:}\quad \bar{r}_{i,t,u}^{(f)} \leq  \delta_{i,t} \min\left\{\hat{r}_{i,t}^{(f)},\bar{r}_{i,t}^{(b)}\right\} x_{i,t,u}, \forall t \in\mathcal{T},\\
\forall i \in\mathcal{I},\forall u \in\mathcal{U}&&\\
\text{C4:}\quad  \bar{r}_{i,t}^{(b)} \leq  \delta_{i,t}\hat{r}_{i,t}^{(b)} y_{i,t} \forall t \in\mathcal{T},\forall i \in\mathcal{I}&\\
\text{C5:}\quad  \alpha_{i,t}, \delta_{i,t}  \in \left\{0,1\right\} \forall t \in\mathcal{T},\forall i \in\mathcal{I}&&\\
\end{aligned}
\end{equation*}
The objective function in \eref{eq:CentralFormula1} aims to minimize both the deployment cost and reconfiguration costs. 

The first term in the objective represents the deployment cost calculated by the total number of deployed extenders. The second term represents the reconfiguration cost and calculated as a function of the number of extender displacement. The first constraint C1 defines the reconfiguration cost as a function of the difference between each two time successive deployments. Thus allows the optimizer to pick the solution that can satisfy all the time horizon demands or requires a small number of changes in the locations of deployed elements. The demand satisfaction of each user at every time instant is captured by the constraint in C2. The set of constraints in C3-C4 are used to calculate the user's throughput (used in C2) as a function of the actual back-haul throughput, its maximum achievable value, and the MAC decisions captured by $ y_{i,t} $ that represent the ratio of AP's resources devoted to extender $ i $. The last constraint in C5 ensures that the two decision variables are binary. Unlike the plethora of Wi-Fi deployment approaches that focused only on optimization techniques to solve similar formulations, we focus also on how network metrics such as throughput can be obtained at minimal cost through learning, and how to leverage the previous decisions to derive future recommendations.

{The computational hardness property of the above defined problem is provided by the following Lemma.}

{ \textbf{Lemma 1.} }\textit{Dynamic Location optimization in WMNs possess the non-deterministic polynomial-time hardness (NP-hard) property.}

{The proof of Lemma is given in Appendix A.}

Hence, below we present a case based heuristic algorithm with semi-supervised learning to achieve a near-optimal deployment of extenders.

\section{AI-Driven Self-Deployment}
A key component of self-deployment is the autonomy, in which the network can optimize the extender location without manual troubleshooting or instructions from operator help desk. To that end, AI is adopted and models the network as two main elements: 1) environment and 2) intelligent agent. The former refers to wireless systems and other objects including neighboring networks, either coordinated or uncoordinated, indoor obstacles and backbone network, among others. The intelligent agent perceives the environment through a sequence of sensing, reasoning and acting in order to build its own knowledge and use it in future actions. Thus, good actions, e.g. satisfies the QoS levels, can be reused directly in future when similar network conditions are sensed, while bad actions, e.g. created coverage holes, will be used to refine the searching strategy of the agent \cite{AIBook}.

The AI can be implemented in different forms such as rule based system (RBS), ontology based system (OBS) and CBR, among others \cite{AISurvey}. The RBS comprises a set of rules with predefined actions created by experts in the network domain. Similarly, OBS applies logic based reasoning for the domain attributes. This logic involves a set of 1) classes: that define a set of objects in the modeled domain; 2) instances: individuals of each class; 3) attributes: properties of each object; and 4) relations: to define relationships between different objects and attributes. 

Unlike CBR, both RBS and OBS require explicit domain knowledge to define the relations between rules and actions or objects. While CBR adopts the system memory to build the knowledge using previous actions and capture their impact on the network. Thus, CBR is well suited for wireless network with none fully observable environment (due to dynamic and uncoordinated deployment, unknown layout plan) and complex operations to be controlled by human-based rules. {However, one of the key issues of CBR are propagation errors through cases. To overcome this issue CBR systems are often used in combination with learning such as reinforcement learning} \cite{cbrrl1} - \cite{cbrrl2}. {The reinforcement learning also benefits from CBR system which significantly speeds up learning of unknown environment and improves its efficiency. Having said that, here we consider CBR in combination with reinforcement learning, where to each problem and solution we assigned the reward value which is determined as fitness value.}

\begin{figure*}[!t]
	\centering
	\includegraphics[width=18cm,trim={3cm 3cm 5cm 5cm},clip]{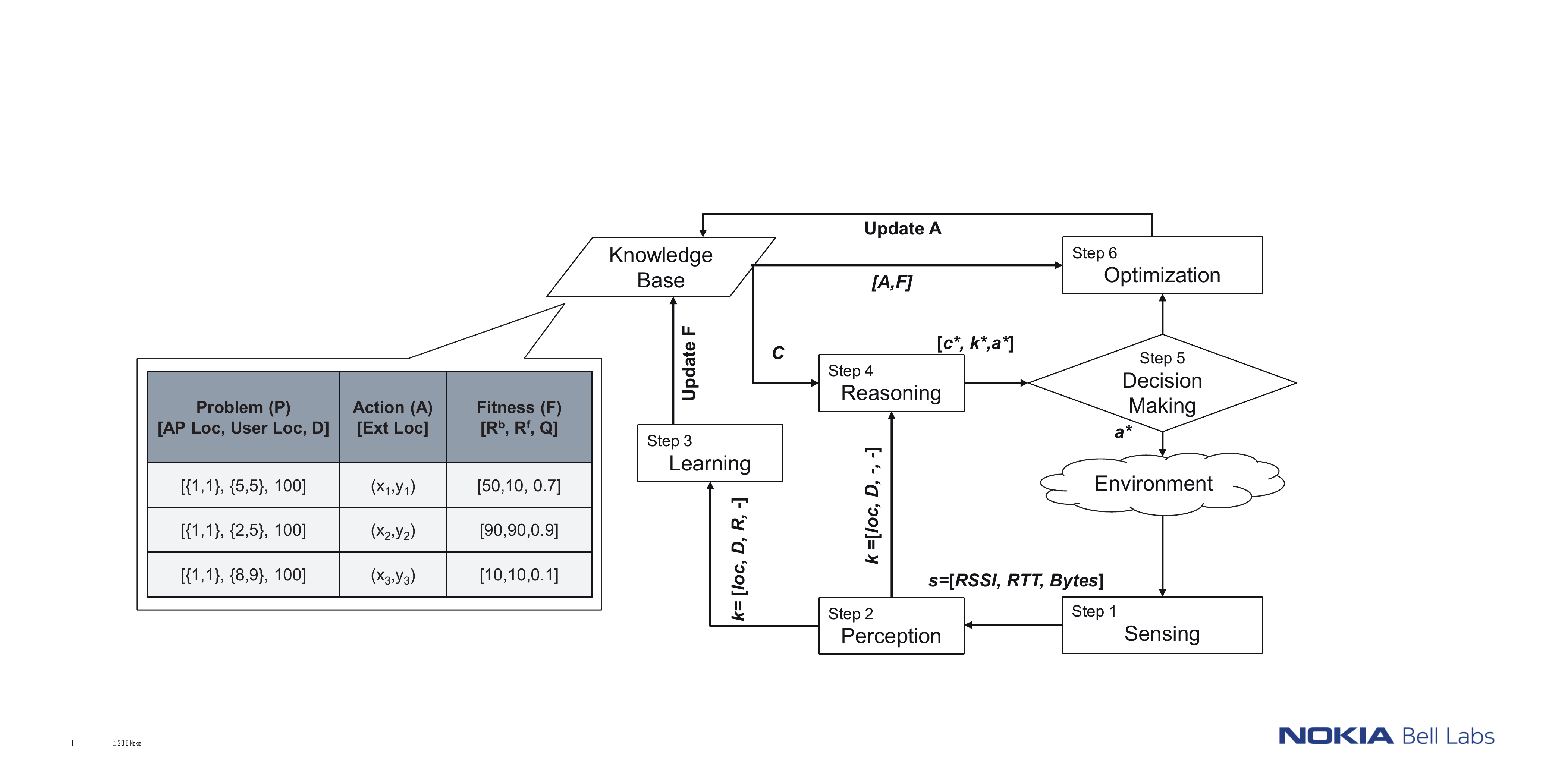}
	\caption{AI-CBR Framework for Self-Deployment.}
	\label{fig:AIFramework}
\end{figure*}	

\subsection{Proposed Framework Overview}
We adopt a CBR AI framework shown in \fref{fig:AIFramework}. The CBR relies fundamentally on a KB that stores previous cases experienced by the self-deployment system, where each case is a triplet of \textit{problem (P)}, \textit{action (A)} and \textit{fitness (F)}. The problem refers to a set of measurements that describes the current situation of the network where a user is unsatisfied. The problem is a vector that contains both the locations and demands of users associated to the extender and the location of AP. For each stored problem, an action can be performed, where action refers to a new position of the extender. {After the action is applied in the environment (i.e. user repositions the extender), the fitness of this action is calculated based on the degree of user demand satisfaction. Accordingly, we denote the fitness for user at location $u$ associated to extender at location $i$ after request $t$, $f_{i,t,u}$ as a ratio between the achievable user throughput, $ r_{i,t,u}$ and the requested user demand, $ D_{t,u}$, i.e.}
\begin{equation}
\label{eq:fitnessFormula}
f_{i,t,u}=\frac{r_{i,t,u}}{D_{t,u}}
\end{equation}
{if $f_{i,t,u}$ is higher than 1, it is set to 1. The fitness value represents the QoS satisfaction degree.}

While the system consistently senses the environment and receives measurements $ s $, the CBR undergoes the following four main stages  \cite{cbr1}, \cite{cbr2}:
\begin{enumerate}
	\item Retrieve the most relevant case, in the knowledge base, to the currently sensed information
	\item Reuse the retrieved case or relative experience to solve the sensed problem
	\item Revise the knowledge base by updating the actions or fitness values of the stored cases
	\item Retain the learned experience (e.g. new case) in the knowledge base to be used in the future
\end{enumerate}

In the following subsections we demonstrate the implementation of every stage using the six main blocks (Step 1-6) in \fref{fig:AIFramework}.  {The main stages of the prosposed framework are summarized in Algorithm} \ref{alg:alg1} {, as well. Unlike the  agent-based approach in} \cite{AIWifi}{, where fully coordinated single-hop Wi-Fi network and the availability of neighbouring environment information are assumed beforehand, our agent continuously senses and learns the} environment to gain experience and generate new actions. The agent enables network intelligence to capture the impact of dynamic neighbouring networks and floor layout while calculating optimal extender's location.

\subsection{Sensing and Perception}
{The first step is sensing the environment which involves collecting the measurements from the user devices, mAP and extenders through TR-98/181 protocol for remote management} \cite{tr181} {or other programmable application interfaces. The collected information contains radio-interface level statistics (e.g used channel width, channel indices, noise level etc.) and user-device level statistics (e.g. RSSI, TxBitRate, RxBitRate, counters for sent and received bytes etc.). The sensing stage collects the data with a certain period $\tau$ in seconds. The perception stage translates the sensed information into performance indicators (i.e. system variables) that identify the network state. This information will allow the agent to perceive the environment, to learn and estimate throughput values for not visited locations and assess the level of user satisfaction.}

{The performance indicators are calculated for each user (user device or extender) based on two successive sensing samples. These indicators include:}
 \begin{itemize}
 \item {Average RSSI - The received signal strength indicator (RSSI) at user location $u$ from sink node (extender) placed at location $i$ after request} $t$, ${RSSI}_{i,t,u}$ {presents a measured received signal strength in dBm of beacon frames received on the channel (i.e. defined as dot11BeaconRssi} \cite{defUtilization}). {RSSI is usually measured during the reception of the physical (PHY) preamble and its value is forwarded to medium access control (MAC) layer in the RXVECTOR} \cite{defUtilization}. {Beacon's RSSI may be averaged over time using a vendor specific smoothing function. In case that the beacon frame is received by means of multiple receive chains, the RSSI is averaged in linear domain over all chains. The valid range of RSSI values  is -100 to 40 dBm} \cite{defUtilization}.
 \item {Average Noise Level can be obtained through the Noise Histogram request/report pair which returns a power histogram measurement of non-IEEE 802.11 noise
power by sampling the channel when virtual carrier sense indicates idle and the STA is neither transmitting nor receiving a frame. This value is denoted as average noise plus interference power indicator (ANIPI)  and its value is contained in Noise histogram report} \cite{defUtilization}.
\item {E2E User Throughput is calculated by using transmitted and received bytes counter values such it is described in section II-A.}
 \end{itemize}

The output of the sensing and perception stage will be used to trigger each of the following stages as discussed below. 


\subsection{Reasoning}
{The perception output will allow the framework to detect if the network is experiencing problems such as unsatisfied user. As an example, when the measured throughput by the perception is lower than the user's demand (i.e. $f<1$), the framework will trigger the reasoning module to compare the current situation (i.e. perception output) with the previously experienced situations (i.e. problem). Reasoning is performed to retrieve the most relevant case from the knowledge base and reuse the corresponding action to solve the current problem. In particular, the current perceived problem $\bar{p}$ is compared to all the stored problems $ P $ in the knowledge base to calculate a matching factor. The action and fitness of the most matching case, denoted by $ a^* $ and $ f^* $ (i.e. corresponding to a single raw in KB), are checked afterwards in the decision making step to determine if they can be reused}. The index $ k^* $ of the most relevant case $ c^* $ is calculated as follows:
\begin{equation}
\label{eq:matchFormula}
\mathbf{k^*}=\underset{{k\in K}}{\text{argmin}}  \quad\,\,\, 
\left\{\sqrt{\sum_{j\in J}(p_{k,j}-\bar{p}_j)^2} \right\}
\end{equation}
where $ K $ is the set of stored cases in the knowledge base and $ {p}_{k,j} $ is the $ j^{th} $ entry of the $ k^{th} $ stored problem. 

{Besides deterministic reasoning other approaches based on probabilistic reasoning (belief network) may be used, but that is out of the scope of this work}.

\subsection{Decision Making}
The decision making step has to determine whether the action of the most matching case can be reused and pushed to the user or a new action has to be recalculated. 

{The decision-making step will check both the calculated matching factor given by Euclidean distance in \eref{eq:matchFormula} and the fitness value given by \eref{eq:fitnessFormula} of the retrieved case.} For instance, the decision making returns true, and applies the action, if both the matching factor is below a maximum threshold, denoted by $ \hat{M} $ while the fitness value $ f^* $ is above a minimal level $ \hat{F} $. If this condition is not met, then either no matching cases exist in the knowledge base and thus a new case must be retained, or the best matching case has a suboptimal action that must be revised.
		
\subsection{Optimization}
The optimization step aims to calculate a new action that will be retained in the knowledge base. In principle, existing actions in the knowledge base will be used to guide the search direction. Two main strategies are followed while creating the new action: exploitation and exploration. The former stands for greedily {optimizing} the network metrics within a limited search space that is assumed to be promising. On the contrary, the exploration tries to discover new search spaces that can lead to more promising solutions than the currently exploited solution set. {The exploration and exploitation strategy} \cite{Mc}, \cite{Th}, \cite{Haykin}{ is decided by policy which is described below. The policy is controlled by an exploration factor whose value is learnt and adjusted with each new action. Although, there are a lot of undirected exploration techniques proposed in literature (such as Random Exploration, Semi-Uniform Distributed Exploration and Boltzmann Distributed Exploration) which randomly explore the environment without consideration the previous history of the learning process, here we apply a directed exploration. It is worth to stress once more, the repositioning of extender involves the user and has a high cost regarding service disruption due to device re-association time which should be minimized. Hence, with this in mind, the random exploration deems unacceptable in this case.}

\subsubsection{Objective}
The first problem to be optimized is the user end-to-end throughput at user devices. While such throughput depends on both the back-haul and front-haul throughput values at the extender, a minimum operator is applied as depicted in the below formulation as follows:
\begin{equation}
\label{eq:AIFormula1}
\left\{
\begin{array}{ll}
\underset{\mathbf{\delta}}{\text{maximize}}  \quad\,\,\, 
 F_{R} = \sum_{\forall i \in I}  {\delta_{i,t}} {\text{min}} \left\{(\hat{r}_{i,t}^{(b)},\hat{r}_{i,t}^{(f)} ) \right\}\\
\text{subject to:}\\
\text{C1:}\quad \sum_i^I \delta_{i,t}  \leq N;~\forall t \in\mathcal{T}\\
\text{C2:}\quad \delta_{i,t}  \in \left\{0,1\right\};~\forall t \in\mathcal{T},~\forall i \in\mathcal{I}
\end{array}
\right.
\end{equation}

The objective function in \eref{eq:AIFormula1} represents the end-user achievable throughput by each deployed extender after request $ t $. It has to be noted that the optimization step adopts estimated values whose accuracies are improved by the learning stage discussed later.

The constraint $C2$ defines the extender location as a binary decision variable, while constraint $ C1 $ bounds the number of selected locations to the $ N $ deployed extenders. 

\subsubsection{Exploration and Exploitation Policy}
The trade-off between exploration and exploitation is very challenging. The exploration strategy enables the optimizer to try different regions where throughput is assumed to be low (e.g. far rooms from the AP but also far from the hidden node neighbor) in order to maximize the acquired knowledge. This exploration is achieved by assigning low fitness values to the previously visited locations, or the positions in their proximity, and vice verse. On the other hand,  the exploitation limits the optimizer to search in a small region  (e.g. in the same room) while no throughput
improvements are observed at the user device.
To that end, the optimizer leverages the saved actions in the knowledge base (i.e. previous visited locations) and calculates their distance to the candidate locations as follows:
\begin{equation}
\label{eq:AIFormula2}
\left\{
\begin{array}{ll}
\underset{\mathbf{\delta}}{\text{maximize}} \quad\,\,\, 
  F_{E}=
\underset{\forall k \in K}{\text{min}} \left\{ {\delta_{i,t}} (\log_{10} \zeta_{k,i})^\omega 
\right\}\\
\text{subject to: C1 - C2}
\end{array}
\right.
\end{equation}		
In the above expression, $ \zeta_{k,i} $ depicts the distance between the candidate location $ i $ and each of the saved locations $ k \in K$ in the knowledge base. The logarithmic function is selected to avoid very far locations from the previously visited. Thus, provide a gradual exploration that does not overestimate such far locations. In this, work, we adopt an exploration factor $ \omega $ that control the degree of exploration of the logarithmic function. Where at low values such as $ \omega =0.1$ the near locations are assigned very high values, which provides limited exploration, i.e. exploitation. On the contrary, high values such as $ \omega =1$ provide steeper decay with the distance, and thus assigns higher fitness to the far locations, not visited before, resulting in wider-scale exploration. The principle of exploration factor is illustrated in \fref{fig:learn}(c). The minimum operator in the objective function selects only one entry from the knowledge base to evaluate the exploration fitness of the candidate locations. 

\subsubsection{Action Generation}
The final action comprises one location that balances both the exploitation and exploration fitness values. Herein, we adopt the multiplication of both values to represent the overall fitness as given by
\begin{equation}
\label{eq:AIFormula3}
\left\{
\begin{array}{ll}
\underset{\mathbf{\delta}}{\text{maximize}} \quad\,\,\, 
\left\{ F_{R} \times F_{E}
\right\}\\
\text{subject to: C1, C2}
\end{array}
\right.
\end{equation}	
Thus, solutions that are far from those previously visited and expected to have high throughput at the end user will have the maximum fitness.

\subsection{Learning}
{The last stage, learning, involves revising and retaining the entries in the knowledge base, and adapting the threshold values and parameters in the other stages based on the measurements. Per each iteration, the learning stage updates the fitness value as in} \eqref{eq:fitnessFormula} {for a certain user at location $u$, and associated to extender at location $i$. While autonomous operation of the AI framework is paramount, yet the user is involved by repositioning the extenders, semi-supervised learning is adopted to learn: 1) system variables (i.e. throughput variables) and 2) exploration factor.}

{Supervised learning techniques are not applicable due to lack of full knowledge about the environment (i.e. floor plan and neighbouring traffic). Thus, in our case, the learning stage is designed to improve the accuracy of system variables by leveraging previously learnt values.}

\subsubsection{Throughput Values}
The throughput estimation is very challenging due to: 1) the dependency of the front-haul on the back-haul value, and 2) the existence of neighbours that can cause hidden nodes. The first necessitates learning the front-haul values only when the back-haul is maximized, while the latter makes the distance-based throughput estimation inaccurate as more interference can be encountered at while placing the extender in the direction of interfering neighbour.

{In essence, the learning stage has to utilize the previously measured throughput values at different locations in order to improve the estimation in other undiscovered locations. After each request $t$, the set of available measurements from the previous $ k \in K $ actions is used as labelled data to estimate the throughput in other not visited locations.}

\begin{itemize}
\item Learned Backhaul Throughput: Semi-Supervised support vector machines (S3VM) \cite{S3VM} are therefore used to update the estimated back-haul throughput $ \hat{r}_{i,t}^{(b)} $ at location $ i $ as a function of: distance based estimated throughput at the current and previously visited locations, and the measured value of the latter as given by
\begin{equation}\label{eq:AIFormula4}
	\hat{r}_{i,t}^{(b)} = \mathcal{F}(\hat{r}_{i,t}^{(b)}, \hat{r}_{k,t}^{(b)}, \bar{r}_{k,t}^{(b)})  \ \ \forall k\in K,  \forall i\in I.
\end{equation}
{Each of the measured throughput values is used to define two regions: between location $ k $ and the AP (i.e. Region 1), and between the location $ k $ and the connected users (i.e. Region 2) as illustrated in} \fref{fig:learn}(b). {According to that,}  $\mathcal{F}$ {has the following definition}
\begin{equation}\label{eq:AIFormula5}
\hat{r}_{i,t}^{(b)}=\left\{
\begin{array}{ll}
 \bar{r}_{k,t}^{(b)} & i \in{\text{Region 1}}\\
\frac{1}{\delta d} \times \bar{r}_{k,t}^{(b)} \ \ \forall k\in K,  \forall i\in I & i \in{ \text{Region 2}},
\end{array}
\right.
\end{equation}
{where} $\delta d$ {is the difference in the distance between locations} $ i $ and $ k $. {While the throughput in the Region 1 remains the same as the measured value, the values decreases gradually in the Region 2. As more measurements become available, the regions are redefined to obtain non-overlapping boundaries using the S3VM in} \cite{S3VM}. {The distance based throughput is used to classify the location in one of the throughput regions. Initial values of estimated throughput variables are calculated as it is mentioned in \sref{sec:netmod}.}
	
\item {Learned Front-haul Throughput: Follows the same procedure as the back-haul except that the gradually decreasing function is applied in the former region (i.e. between AP and extender). The main challenge is to decouple the impact of back-haul on the front-haul throughput calculation. As such, a region is defined in which the network guarantees that the reported front-haul throughput is only due to the channel conditions between the extender and the user, and not due to poor back-haul link. As such, this region is defined as the area between the extender and user locations in which the back-haul throughput satisfies the demand or surpasses the total front-haul. Thus, the updated estimated front-haul throughput} $\hat{r}_{i,t}^{(f)}$  at location $ i $ {is a function of: 1) distance based estimated throughput at the current and previously visited locations, and 2) the measured value of the latter as given by}
\begin{equation}\label{eq:AIFormulaFT}
	\hat{r}_{i,t}^{(f)} = \mathcal{F}(\hat{r}_{i,t}^{(f)}, \hat{r}_{k,t}^{(f)}, \bar{r}_{k,t}^{(f)})  \ \ \forall k\in K,  \forall i\in I.
\end{equation}
{where} $\mathcal{F}$ {is defined in \eref{eq:AIFormula5}, while Region 1 is defined as a region between location $i$ and location $k$, and Region 2 is defined as region between location $k$ and location $n$, where is $n>k$.}
\end{itemize}

\subsubsection{Exploration Factor}
\begin{figure}[h]
	\centering
	\subfigure[Learning backhaul throughput: illustration of back-haul throughput in \eref{eq:AIFormula4}]
	{\includegraphics[scale=0.34]{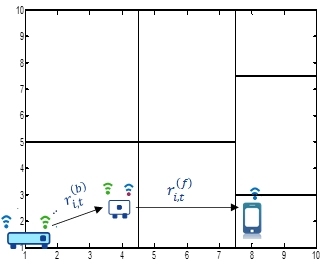}}\hspace{0.8em}\vspace{-0.3em}%
	\label{fig:learnA}
	\subfigure[Learning backhaul throughput: region definition and estimation in \eref{eq:AIFormula7}.]
	{\includegraphics[scale=0.34]{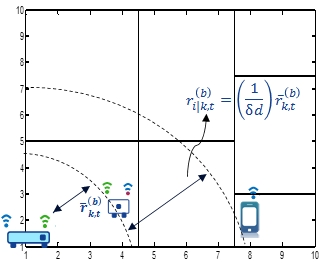}}\vspace{-0.3em}%
	\label{fig:learnB}
	
	\subfigure[Principle of exploration step in \eref{eq:AIFormula2}]{\includegraphics[scale=0.34]{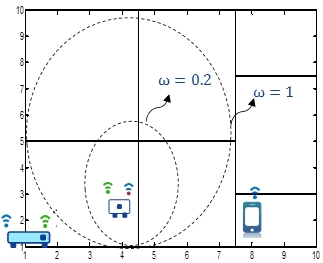}}\hspace{0.8em}\vspace{-0.3em}%
	\label{fig:learnC}
	\subfigure[The adjustement of Exploration Factor]	{\includegraphics[scale=0.34]{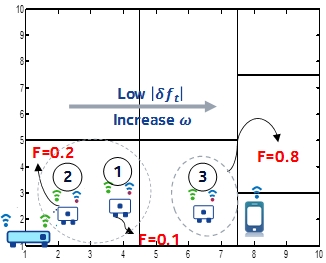}}\vspace{-0.3em}%
	\label{fig:learnD}
	\caption{Illustration of different learning principles.}
\label{fig:learn}
\end{figure}

{The exploration factor is a key of policy to control exploitation and exploration and it controls how far the extender should be placed from the previously visited locations that are deemed suboptimal}. The exploration factor $ \omega $ is thus adapted based on the reported fitness values to guarantee visiting spatially separated regions. As such, the value of $ \omega $ has to be increased, thus higher exploration, in case of similar fitness values obtained by the generated actions, and vice versa. This strategy is depicted as follows:
\begin{equation}
\label{eq:AIFormula7}
\left\{
\begin{array}{ll}
\omega_t = \frac{1}{2}(\omega_{t-1} + \delta \omega_t)~\forall t\in T&(a)\\
\delta \omega_t = 2- e^{|\delta F|}&(b)
\end{array}
\right.
\end{equation}	
In \eref{eq:AIFormula7}(a), the exploration factor is updated after each request $ t $ using the previous value and the step variable $ \delta \omega_t $. The latter is calculated based on the difference in the fitness values of the last two actions. The exponential function in \eref{eq:AIFormula7}(b) results in negative values if the absolute fitness difference $ |\delta F| $ is high which will decrease the value of $ \omega_t $ in \eref{eq:AIFormula7}(a) to limit the exploration, i.e search in limited region. On the other hand, low $ |\delta F| $ results in higher $ \omega $ that maximizes the exploration as illustrated in \fref{fig:learn}(c). It has to be noted that $ \delta F \in [-1,1] $ and thus the calculated value in \eref{eq:AIFormula7}(a) is normalized.


\begin{figure}[!t]
	\centering
	\includegraphics[width=0.8\linewidth,trim={0cm 4cm 0cm 2cm},clip]{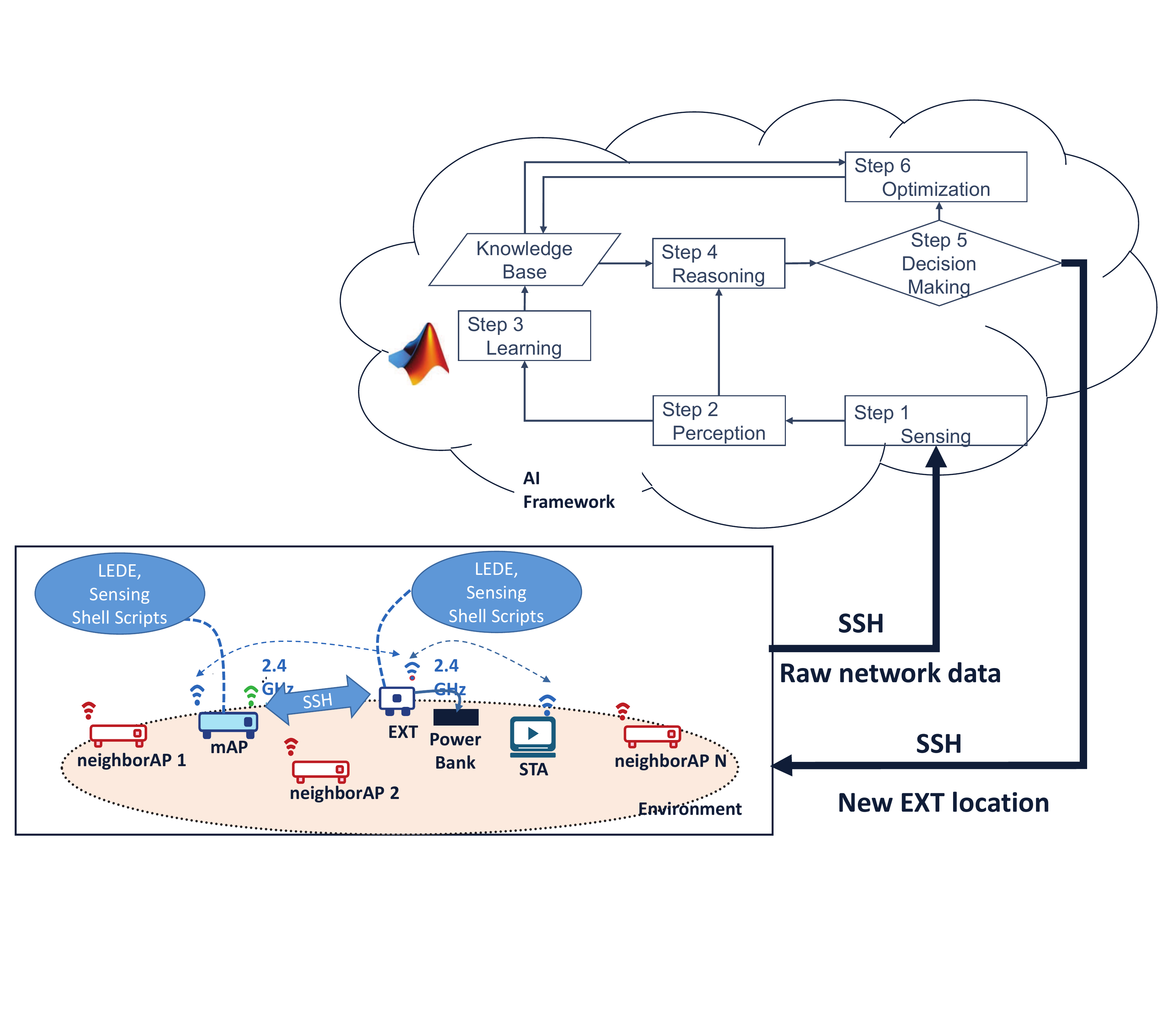}
	\caption{AI Testbed with off-the-shelf devices.}
	\label{fig:testbedLogic}
\end{figure}

\begin{algorithm}
\begin{small}
\caption{ Dynamic Location Optimization}
\label{alg:alg1}
\SetAlgoLined
\SetKwInOut{Input}{Input}
\SetKwInOut{Output}{Output}
\SetKwInOut{Initialization}{Initialization}
\Input {Knowledge Base (Cases, Fitness Values, User Demands, Perception data)\;}
\Output  {Action $a^*$\;}
\textbf{Define:} Max. matching threshold: $\hat{M}$, min. fitness threshold: $\hat{F}$, max. number of repositioning actions $N$\;
\For {each extender $i$ in the network}
{
Update estimated backhaul throughputs using Eq. \ref{eq:AIFormula5}\;
\For {$u \in U$ associated to extender $i$}
{
Update estimated fronthaul throughputs using Eq. \ref{eq:AIFormulaFT}\;
Update fitness value for user at location $u$ associated to extender $i$  using Eq. \ref{eq:fitnessFormula}\;
/* Check if the user demand is satisfied */ \\
\If {$r_{i,t,u}<D_{t,u}$}
{
/*Reasoning*/ \\
Find the best matching case $c^*:=(a^*,f^*)$ in KB using Eq. \ref{eq:matchFormula}\;
/*Decision Making */ \\
\If {$ k^*< \hat{M} \text{  and  } f^* > \hat{F}$}
{
apply action {$a^*$}
}
\Else
{
/*Optimization*/ \\
Adjust exploration factor by Eq. \ref{eq:AIFormula7} and generate new action $a^\prime$ by Eq. \ref{eq:AIFormula3}\;
\textbf{if} $\textit{total number of extender repositioning}<N$

\hspace{0.1cm} apply action {$a^\prime$}

\textbf{end}

}

}

}
}

\end{small}
\end{algorithm}

\section{Performance Evaluation}
\subsection{{Experimental} Setup}
\subsubsection{Testbed Environment}
\begin{figure}[!t]
	\centering
	\subfigure[Environment]	{\includegraphics[width=0.5\linewidth,trim={0cm 0cm 0cm 0cm},clip]{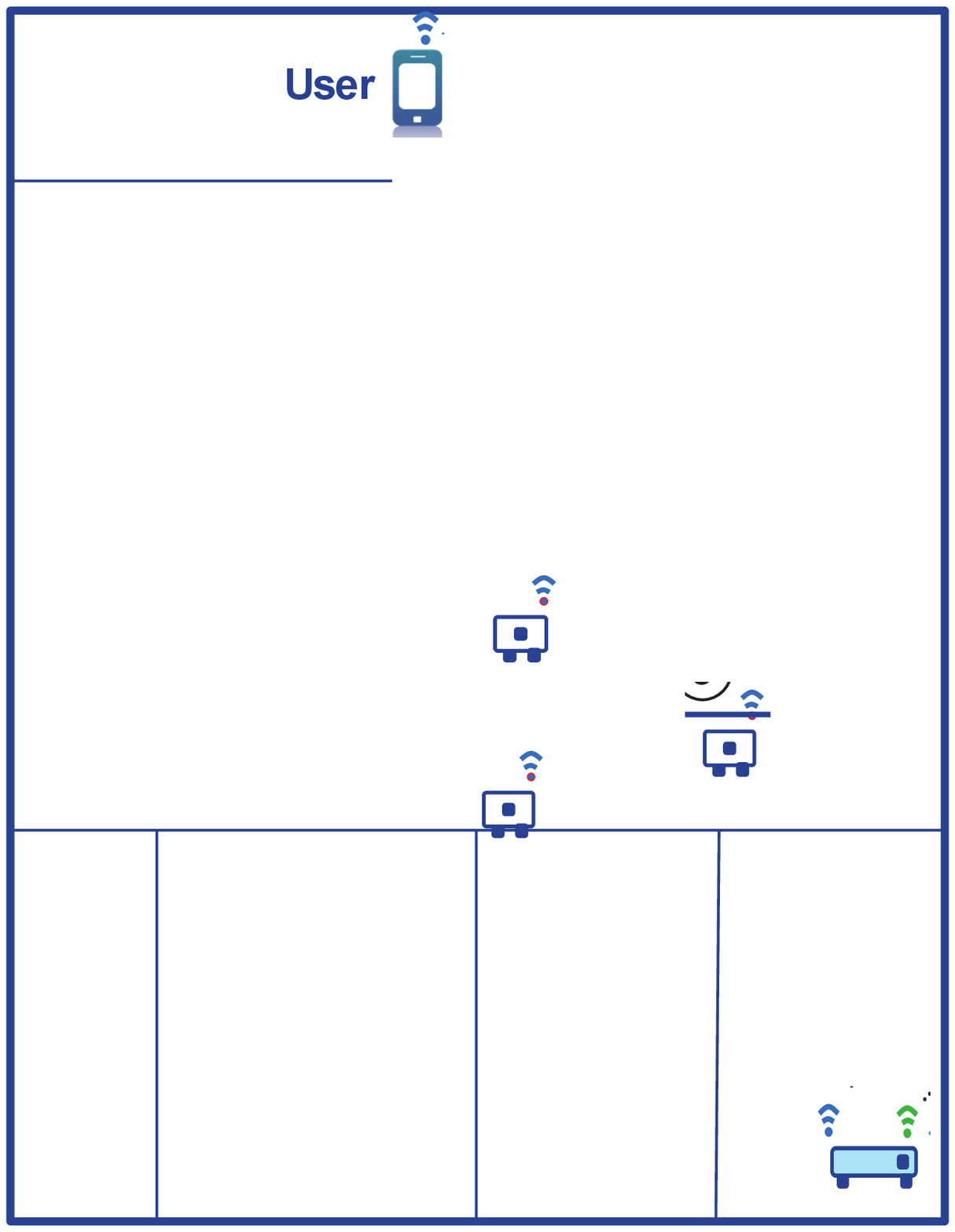}}%
	\subfigure[Throughput Measurements]
	{\includegraphics[width=0.5\linewidth,trim={0cm 0cm 0cm 0cm},clip]{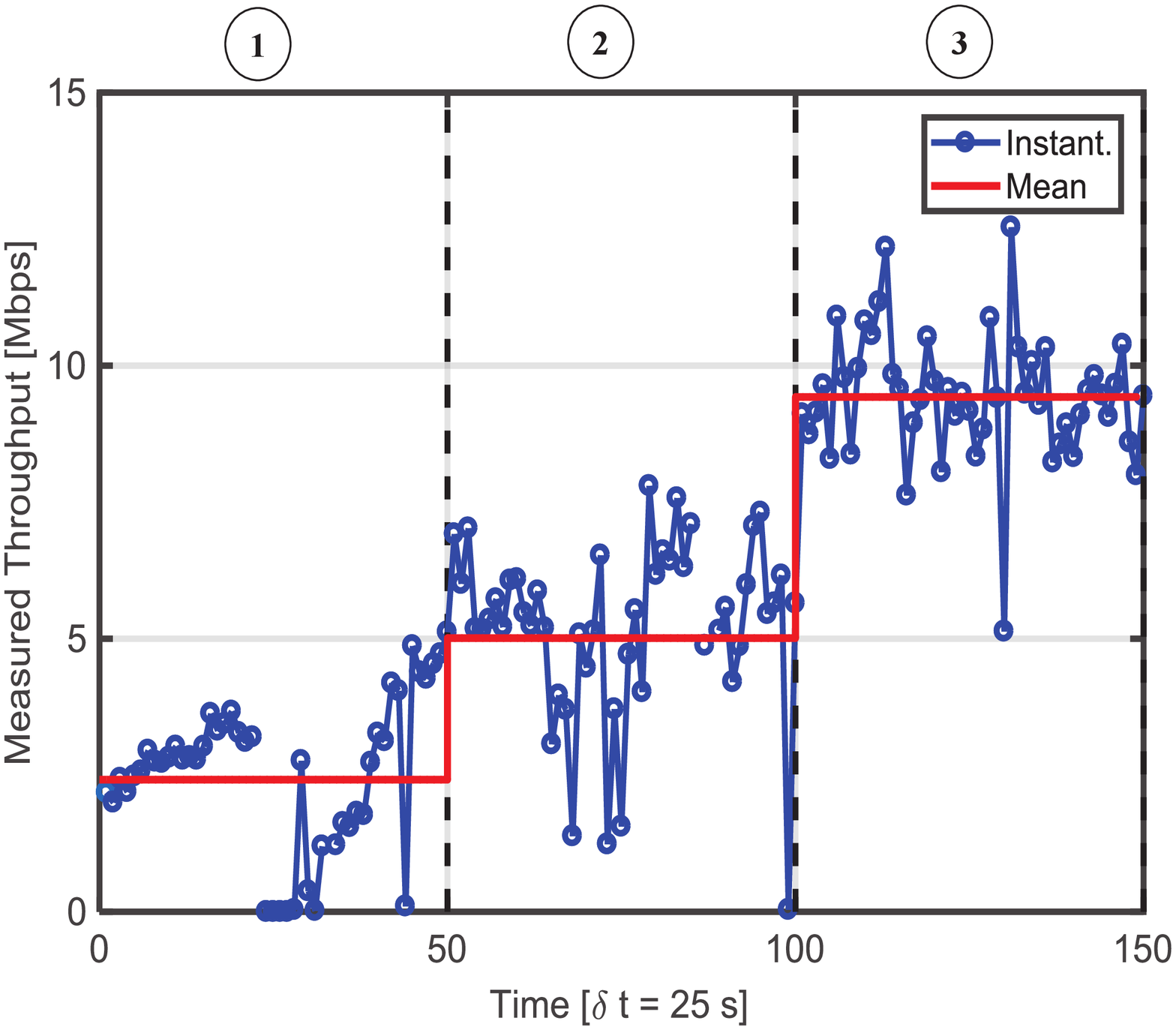}}%
	\caption{Testbed results for coverage problem (midway initial position).}
	\label{fig:Res1}
\end{figure}

The experimental testbed consists of AI framework, a single wired-backhaul AP, i.e. mAP, a single extender (EXT) and an end-user device which is connected to mAP through EXT. AI framework logic is implemented in MATLAB and hosted in the cloud. AI engine has a secure connection to mAP, which is used to collect network parameters and to push change-location notifications to the end users via embedded speaker. For mAP we consider dual-band TP-Link AC1750 AP, whereas EXT is a single-band GL-MT300A {operating} on 2.4 GHz. EXT is configured as a simple repeater of the signal received from mAP and the signal received from the user devices associated to itself. In other words, EXT operates as a station connected to mAP at the side of backhaul link and as an access point operating in infrastructure mode on the side of fronthaul link. Since EXT is battery powered (connected to a power bank via USB), its location is not restricted to the locations where electric plugs are available. We equipped EXT with usb-to-audio adapter and a speaker device, in order to be able to produce audible change-location notifications to the end users. The indoor localization information is assumed available. We installed LEDE images \cite{lede} on both mAP and EXT and by means of shell scripts, periodical network parameters are reported to the cloud hosting the AI framework. mAP and EXT are deployed in an uncoordinated environment whose layout is shown in \fref{fig:Res1}(a) .


\begin{figure}[!t]
	\centering
	\subfigure[mAP location]		{
	\includegraphics[width=0.45\linewidth,trim={0cm 0cm 0cm 0cm},clip]{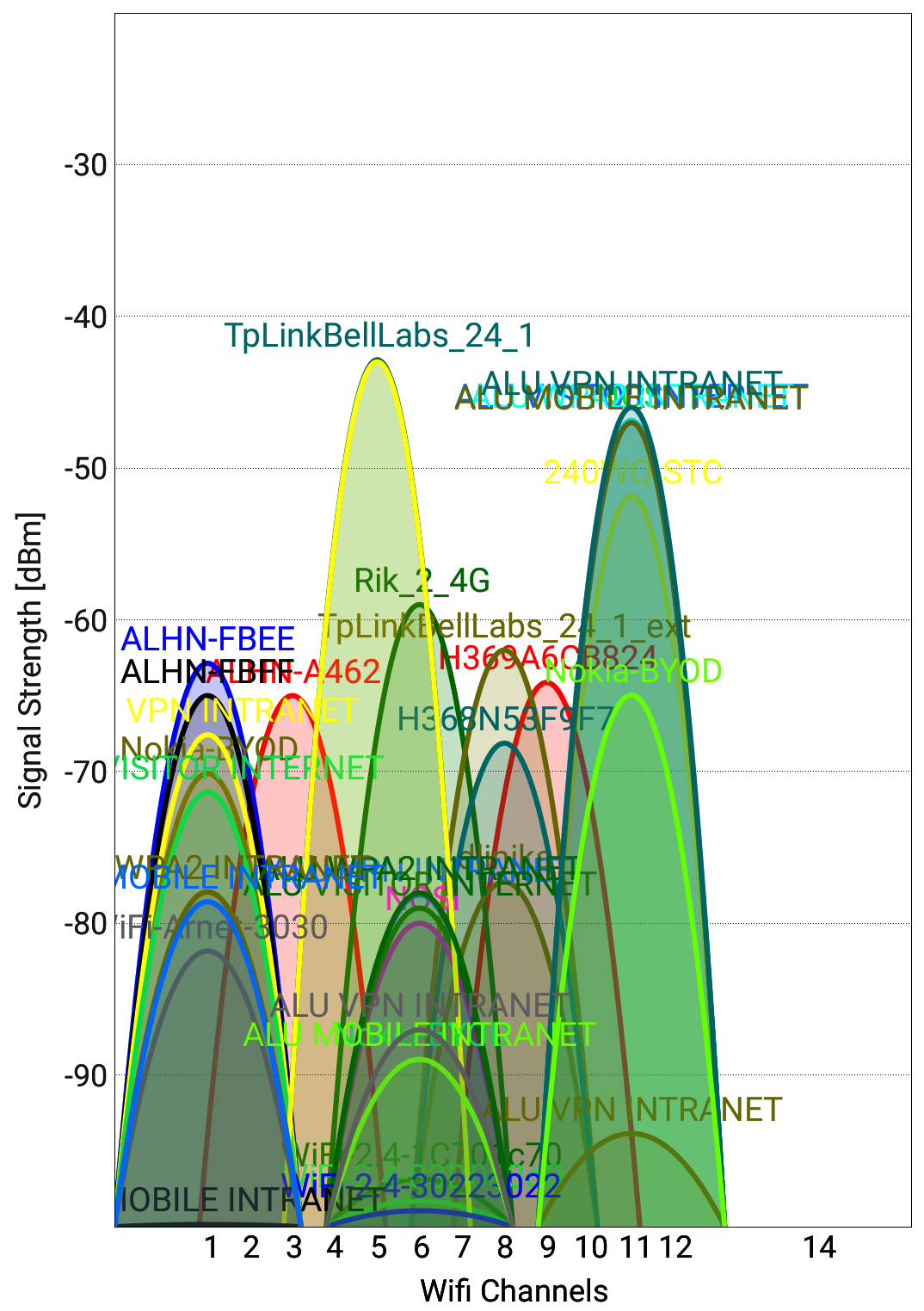}}
	\hspace{0.8em}
	\subfigure[EXT at location 1]
	{\includegraphics[width=0.45\linewidth,trim={0cm 0cm 0cm 0cm},clip]{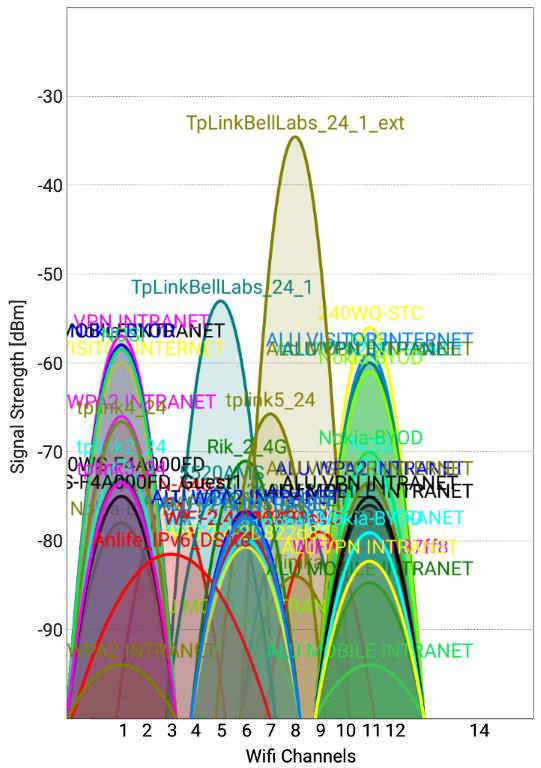}}
	\caption{2.4 GHz wireless spectrum at the locations of mAP and EXT in \fref{fig:Res1}(a) at location 1 (similar density of neighbouring networks is observed at locations 2 and 3 with distance of 2 and 4 meters, respectively, from location 1.}
	\label{fig:scan}
\end{figure}

\subsubsection{Coverage Problem}
We applied the experimental testbed in an enterprise scenario as shown in \fref{fig:Res1}(a). The environment employs a large number of uncoordinated neighboring Wi-Fi APs as illustrated by the scan of the spectrum in \fref{fig:scan}. Due to different received power levels of neighboring APs, exposed and hidden nodes are experienced by the coordinated network on which the AI framework is tested. With such an ultra dense scenario, we manually set the operating channel after manual tuning to minimize contention at the AP. Other parameters are summarized in \tref{table:parameters}.
 The extender is initially placed in the mid-way between the AP and user device. The throughput measurements are reported in \fref{fig:Res1}(b) for the three locations of extender where the location index is shown in circles in \fref{fig:Res1}. The second and third locations are successively calculated by the proposed AI framework and resulted in an end-to-end throughput improvement of $ 300 \% $ at the last location compared to the initial mid-way based one. In particular, The AI-framework  gradually improves the backhaul throughput by placing the extender closer to the AP. The third location, i.e. the optimal one, compensated the high partition attenuation factor of the glass walls between the mAP and EXT by minimizing the length of backhaul link. This placement leverages the open space between the user device and EXT, and thus the fronthaul throughput is not deteriorated by moving the extender away from the user. 

\begin{figure}[!t]
	\centering
	\subfigure[Environment]
	{
		\includegraphics[width=0.5\linewidth,trim={0cm 0cm 0cm 0cm},clip]{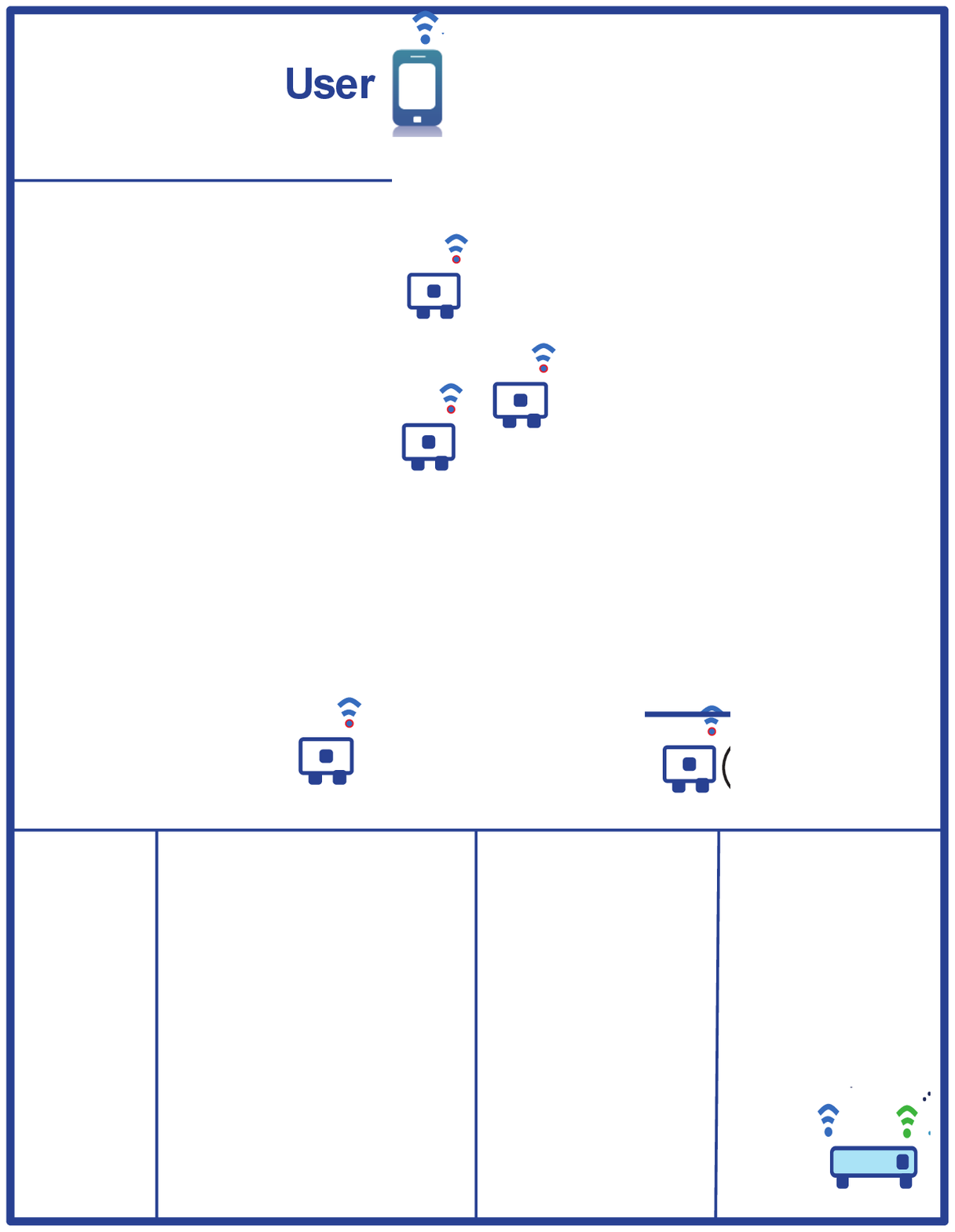}
	}%
	\subfigure[Throughput Measurements]	
	{
		\includegraphics[width=0.5\linewidth,trim={0cm 0cm 0cm 0cm},clip]{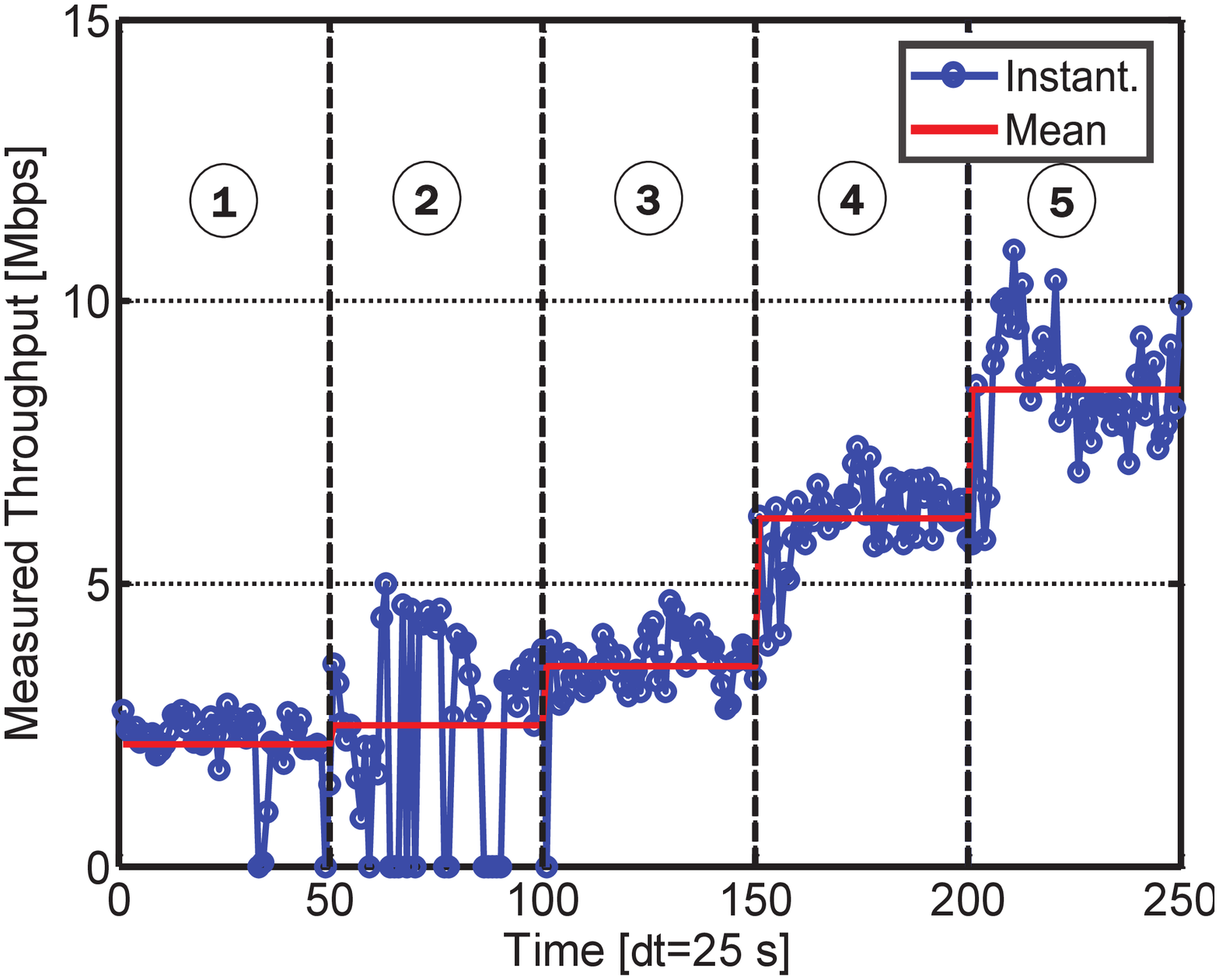}
	}%
	\caption{Testbed results for coverage problem (random initial position).}
	\label{fig:Res2}
\end{figure}

We repeated such coverage problem but with a different initial location in order to assess the convergence of the AI approach as depicted in \fref{fig:Res2}(a). In this case, more locations were recommended before the final optimal location is reached as shown in \fref{fig:Res2}(b). Thus, the cost of learning  is said to be higher, equals to 3, than the previous scenario whose cost of learning was 1. The intermediate locations 2 and 3 in \fref{fig:Res2}(a) demonstrate the exploitation phase where the region surrounding the initial suboptimal location is searched first. However, recommended locations did not result in significant throughput improvements as show in \fref{fig:Res2}(b). Thus, the algorithm attempted to explore the environment and recommended locations 4 and 5 that are farther from the last recommended locations.

\subsubsection{Interference (Hidden Node) Problem}	
\begin{figure}[!t]
	\centering
	\subfigure[Environment]
	{
		\includegraphics[width=0.5\linewidth,trim={0cm 0cm 0cm 0cm},clip]{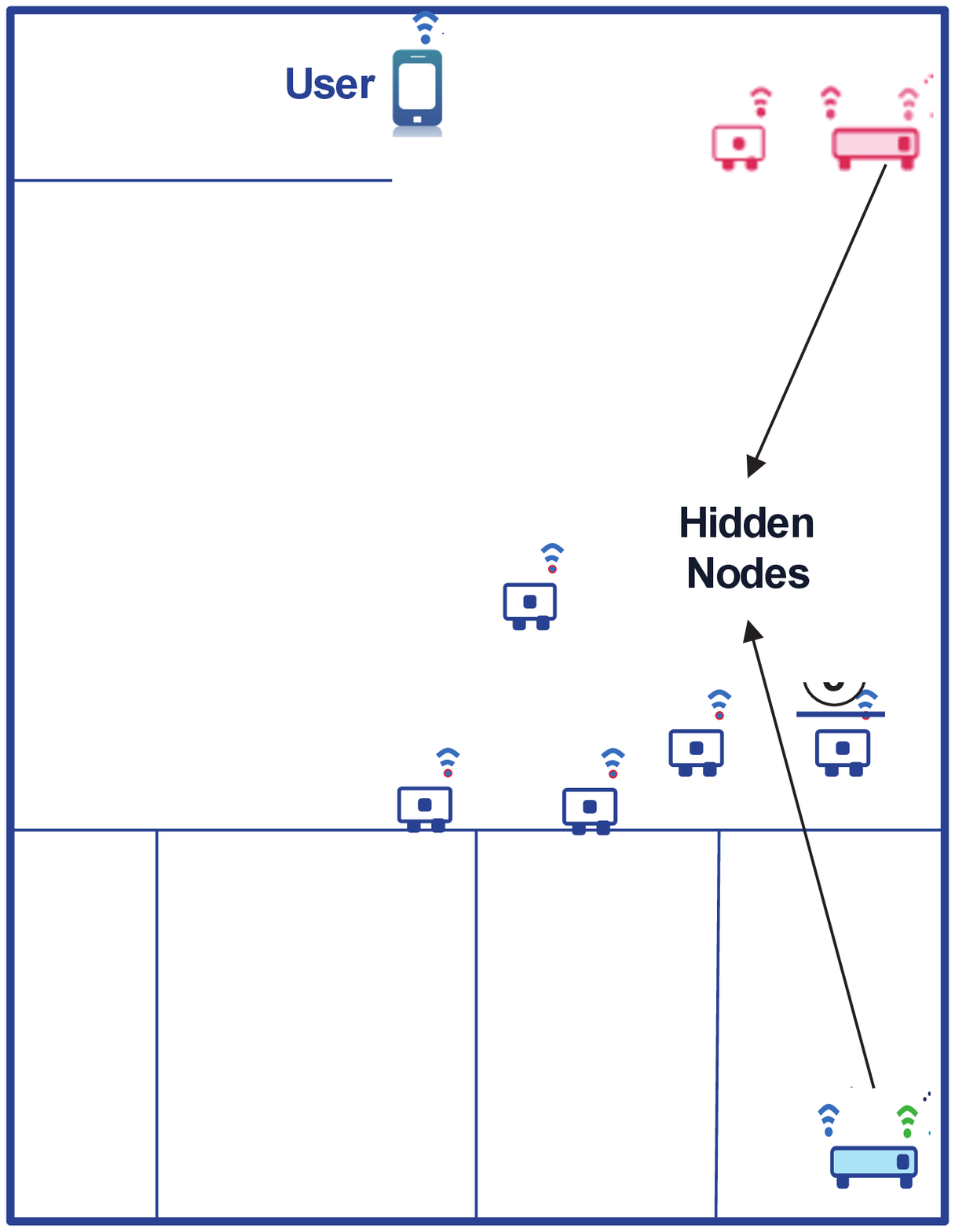}
	}%
	\subfigure[Throughput Measurements]	
	{
		\includegraphics[width=0.5\linewidth,trim={0cm 0cm 0cm 0cm},clip]{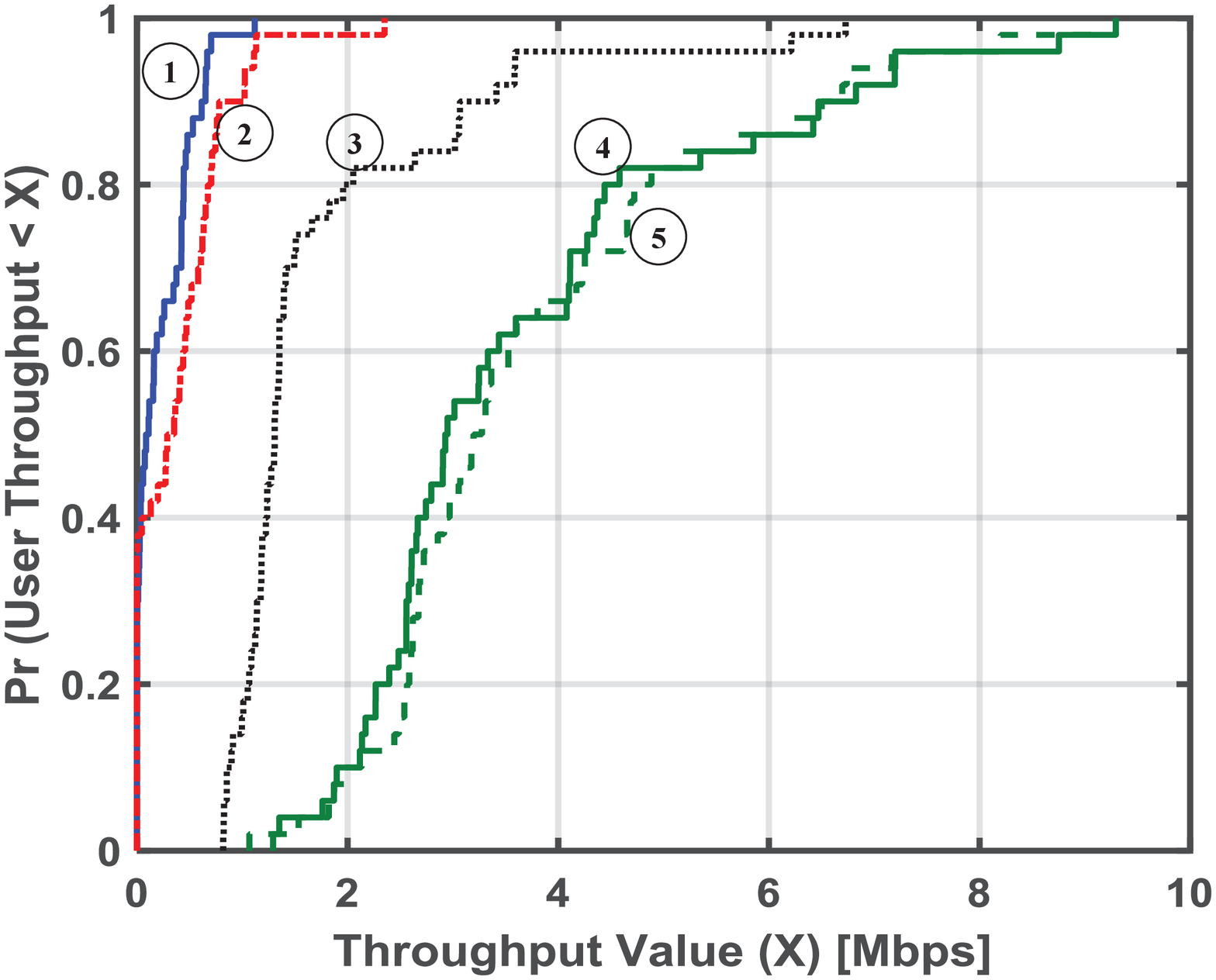}
	}%
	\caption{Testbed results for interference scenario with hidden node problem.}
	\label{fig:Res3}
\end{figure}

Beside the existing unmanaged neighboring networks, an interfering network (i.e. AP, extender and user device) is deployed close to the managed user device and operate at the same channel. The location of such interfering AP is chosen such that it acts as a hidden node to the managed AP and create excessive interference at the BH of the managed extender as shown in \fref{fig:Res3}(a). To demonstrate the worst case interference, the interfering network is working in a saturated traffic mode and hence occupying the shared medium all the time. Such hidden node scenario resulted in very low throughput values in the initial location with a service outage $ 40\% $ of the collected measurements as depicted in \fref{fig:Res3}(b). This is in addition to increasing the cost of learning to 2 although the same initial location as the first scenario was adopted. Comparing the throughput of the final optimal locations, i.e. 4, with the initial mid-way location, i.e. 1, the former resulted in more than $ 11 $ times throughput improvements without any service outages.

\subsubsection{User Experience}
In the above experiments, the user was watching a Full High Definition (FHD) 2K YouTube video while connected to the extender. The user's quality of experience (QoE) can be typically measured as a function of duration of video stops which is a critical parameter reflecting the user satisfaction. Herein, the QoE value is calculated by Mean Opinion Score (MOS) according to the model in \cite{MOSImportant2}, where MOS varies between 1 and 5 corresponding to very poor and excellent service, respectively. In case of suboptimal placement of extender such as mid-way location (i.e. location 1) as in \fref{fig:Res2}(a), the user watched a $ 2 $ minutes video in a duration of $ 3.6 $ minutes. These stops are attributed to the buffer underrun at the user device since the delivered throughput is not high enough to transmit the high quality video content on time. Thus, this stop duration of $ 1.6 $ minutes resulted in a MOS around $1 $ which is not acceptable by end-users. On the contrary, optimizing the location of extender results in zero stops and thus the maximum MOS was achieved.

\subsection{Simulation Setup}
\subsubsection{ns-3 Environment}
To complement experimental results with single-band extender and single user setup, we evaluate the proposed method using the Wi-Fi IEEE 802.11ac compliant simulator ns-3. The extender is modeled as a node that implements two Wi-Fi interfaces, one working in ad-hoc mode, to communicate with the AP nodes, while the other interface is operating in an infrastructure mode to act as an AP for the user station (STA). The two interfaces are operating on two different radios (i.e. dual-radio) both on the $ 5 $ GHz band. All the simulation parameters and values are summarized in \tref{tab:notation}. The maximum achievable gains of the proposed method are computed by adopting exhaustive search in solving the above presented optimization problem. In this section, we further resort to computer simulation to get insights and investigate the effectiveness of AI framework in more complex deployments. 

We compare the proposed framework, referred to as \emph{AI-Driven CBR}, with ($i$) the coverage maximization and delay minimization approaches in \cite{Ext1}, referred to as \emph{Coverage Max.}, and ($ii$) the AP only scenario without extenders.
This is in addition to reporting results for the AP only scenario to illustrate the impact of creating multi-hope networks, due to an extender, on the network performance. The proposed framework will be referred to as \emph{AI-Driven CBR}. The number of reconfiguration requests to reposition the extender is set to $ 5 $ to minimize the burden. Results are reported for $ 50 $ simulation drops where the locations of users are randomly changed. The valuation metrics is summarized below:
\begin{itemize}
	\item Average Throughput: The first metric adopted in this evaluation is the average throughput among all the users served by the same AP, either directly or through an extender. Thus, illustrates the ability of self-deployment to maximize the resource utilization in the network.
	\item Fairness Index: The second metric is the throughput fairness calculated using Jain's fairness index \cite{Jain}. This represents the ability of self-deployment scheme to achieve a uniform QoS satisfaction among the users. 
	\item Service Outage: This metric refers to the percentage of scenarios when the users receive zero throughput. As such, represents the risk of having throughput holes in the environment.
\end{itemize}

\renewcommand{\arraystretch}{.9}
\begin{table}[!t]
\caption{Summary of Parameters for Experiment}
\label{table:parameters}
\centering
\begin{tabularx}{.85\columnwidth}{lX}
\hline
Parameter & Value \\ \hline	
Transmit power (AP and extender) & $20$\,dBm \\ 
Bandwidth & $20$\,MHz\\
Frequency Band & $2.4$\, GHz \\
YouTube Video Quality & $2K$\, \\
Transmission Rate & Fixed: MCS $ 5 $ \\
Number of radios in extender & $ 1 $ \\
\hline
\end{tabularx}
\end{table}	

\renewcommand{\arraystretch}{.9}
\begin{table}[!t]
	\caption{Summary of Simulation Parameters}
	\label{tab:notation}
	\centering
	\begin{tabularx}{.85\columnwidth}{lX}
		\hline
		Parameter & Value \\ \hline	
		Transmit power (AP and extender) & $20$\,dBm \\ 
		Bandwidth & $80$\,MHz\\
		Frequency Band & $5$\, GHz \\
		Demand & $\left\{100,150\right\}$\,Mbps \\
		Path Loss Model & IEEE 802.11ax \\
		Wall Loss & $ 10 $\, dB \\
		Receiver Sensitivity & $ -83 $ dBm \\
		STA Locations & $ U[0,10]$ \\
		Rate Adaptation & Minstrel \\
		Number of radios in extender & $ 2 $ \\
		\hline
	\end{tabularx}
\end{table}	

\subsubsection{Results}
\begin{figure}[!t]
	\centering
	\subfigure[2 apartments scenario]
	{
		\includegraphics[width=0.6\textwidth,trim={1cm 1cm 1cm 0cm},clip]{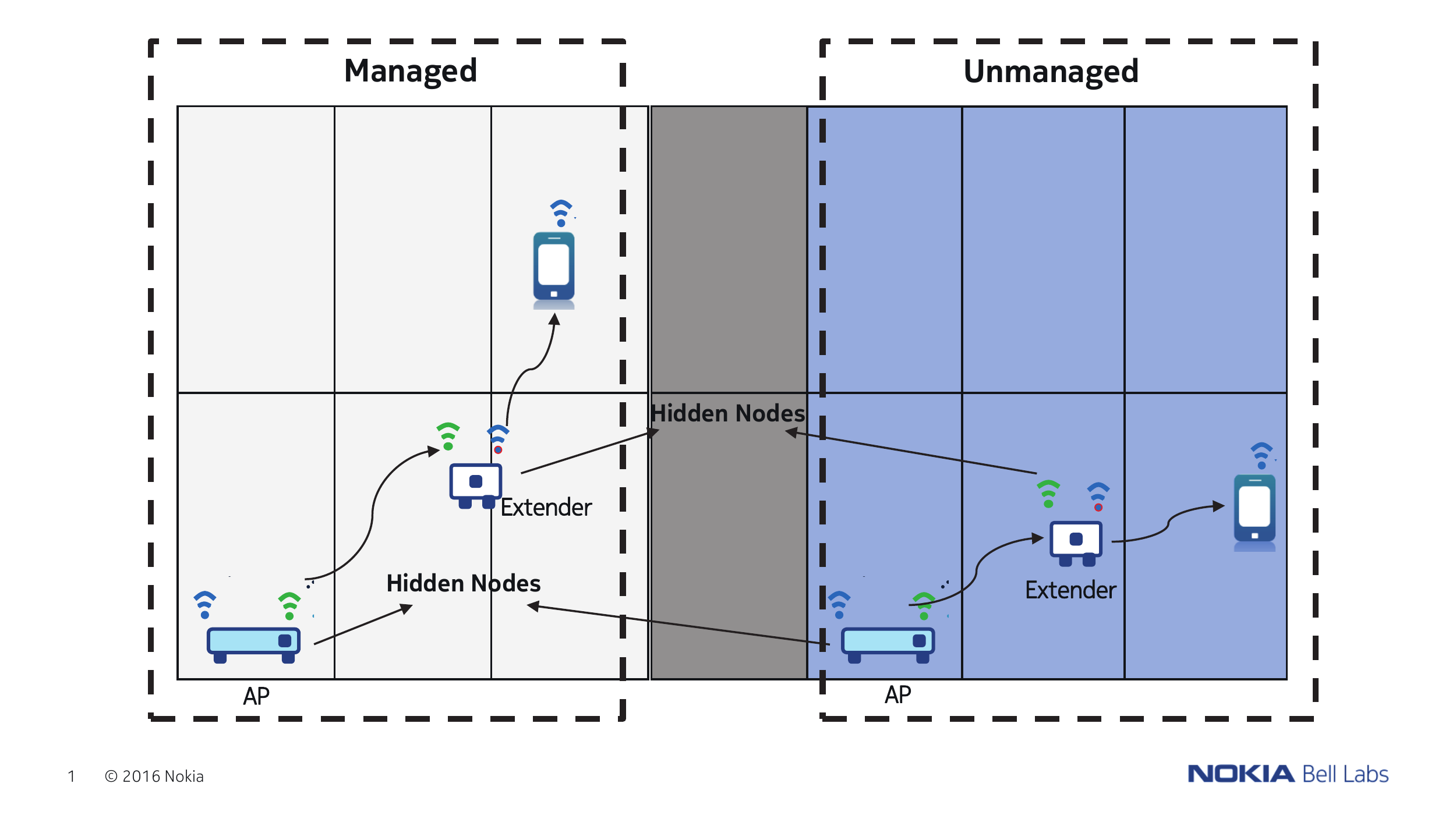}
	}
		\label{fig:scenario1}
	\subfigure[1 Floor, 10 apartments (1STA Each) scenario]
	{
		\includegraphics[width=0.6\textwidth]{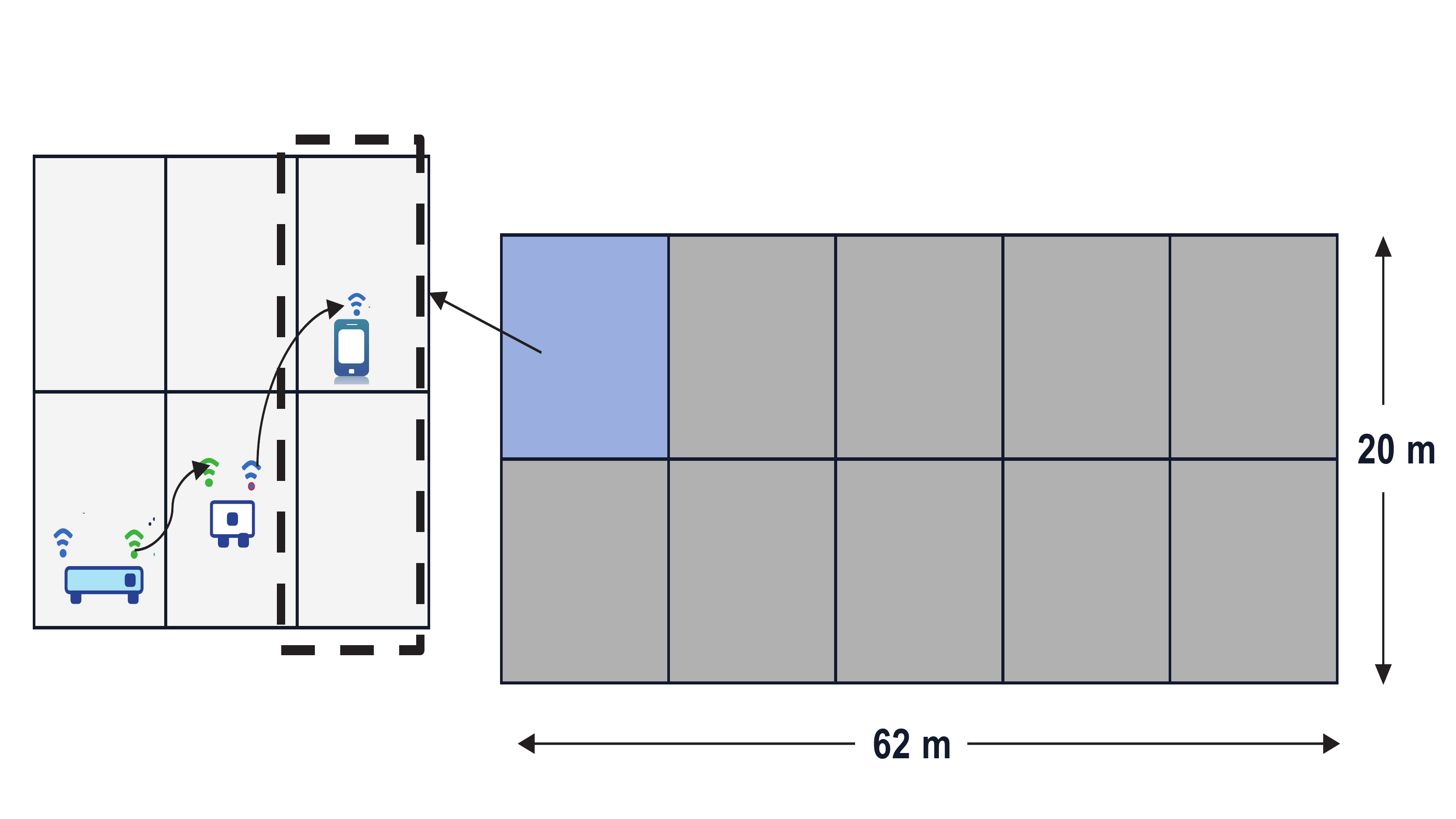}
	}
	\caption{Example of simulated uncoordinated Wi-Fi indoor deployment.}
	\label{fig:SimScenarios}
\end{figure}

We simulate two isolated scenarios in which 1 apartment is considered without neighbors, and two uncoordinated scenarios with unmanaged neighboring apartments. The apartment is a square with side length of $ 10 $ m and comprises 6 rooms as illustrated in \fref{fig:SimScenarios}.

\begin{figure*}[!t]
	\centering
	\subfigure[100 Mbps 2 Users]
	{
		\includegraphics[width=0.48\textwidth]{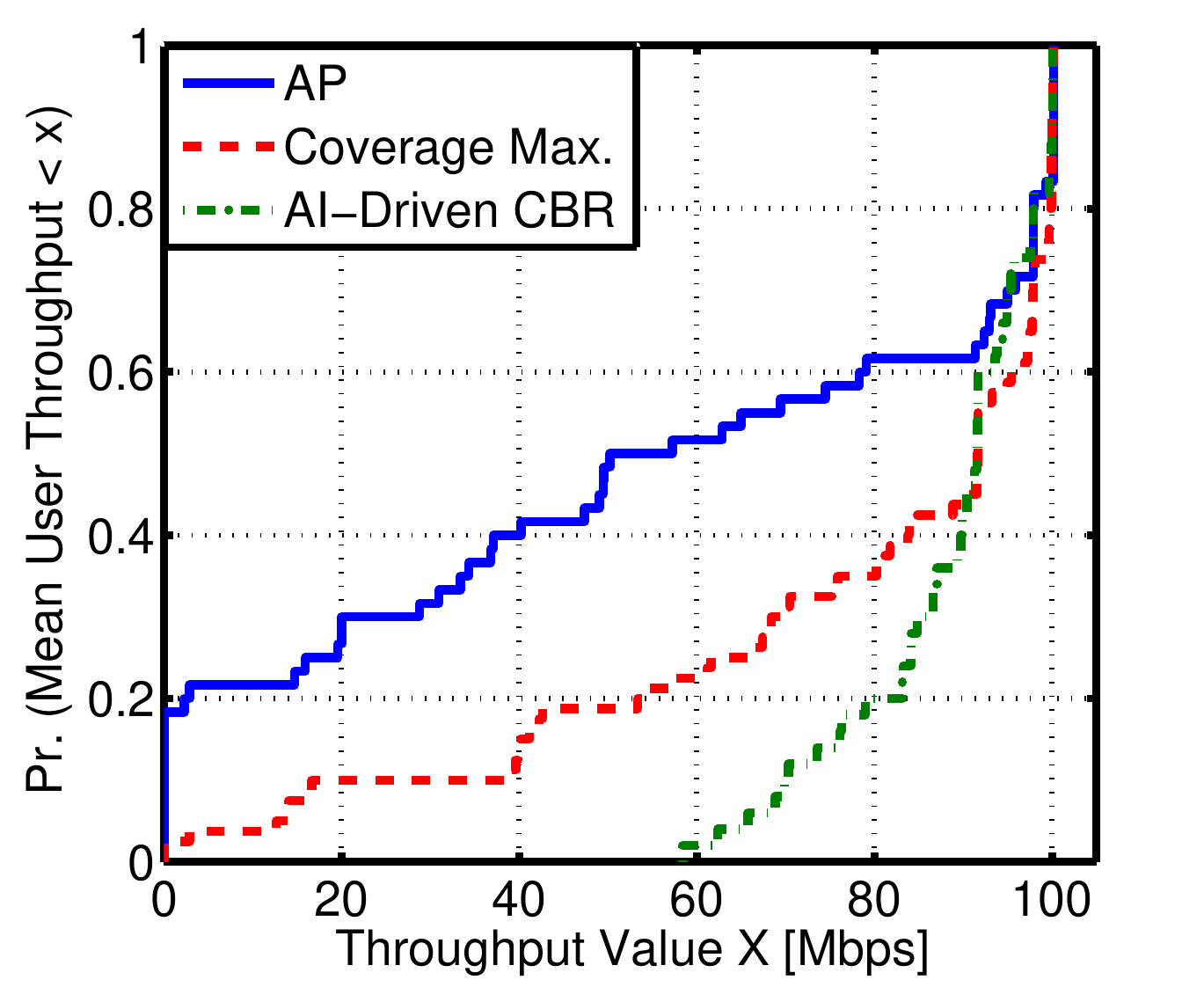}
	}
	\subfigure[150 Mbps 2 Users]
	{
		\includegraphics[width=0.48\textwidth]{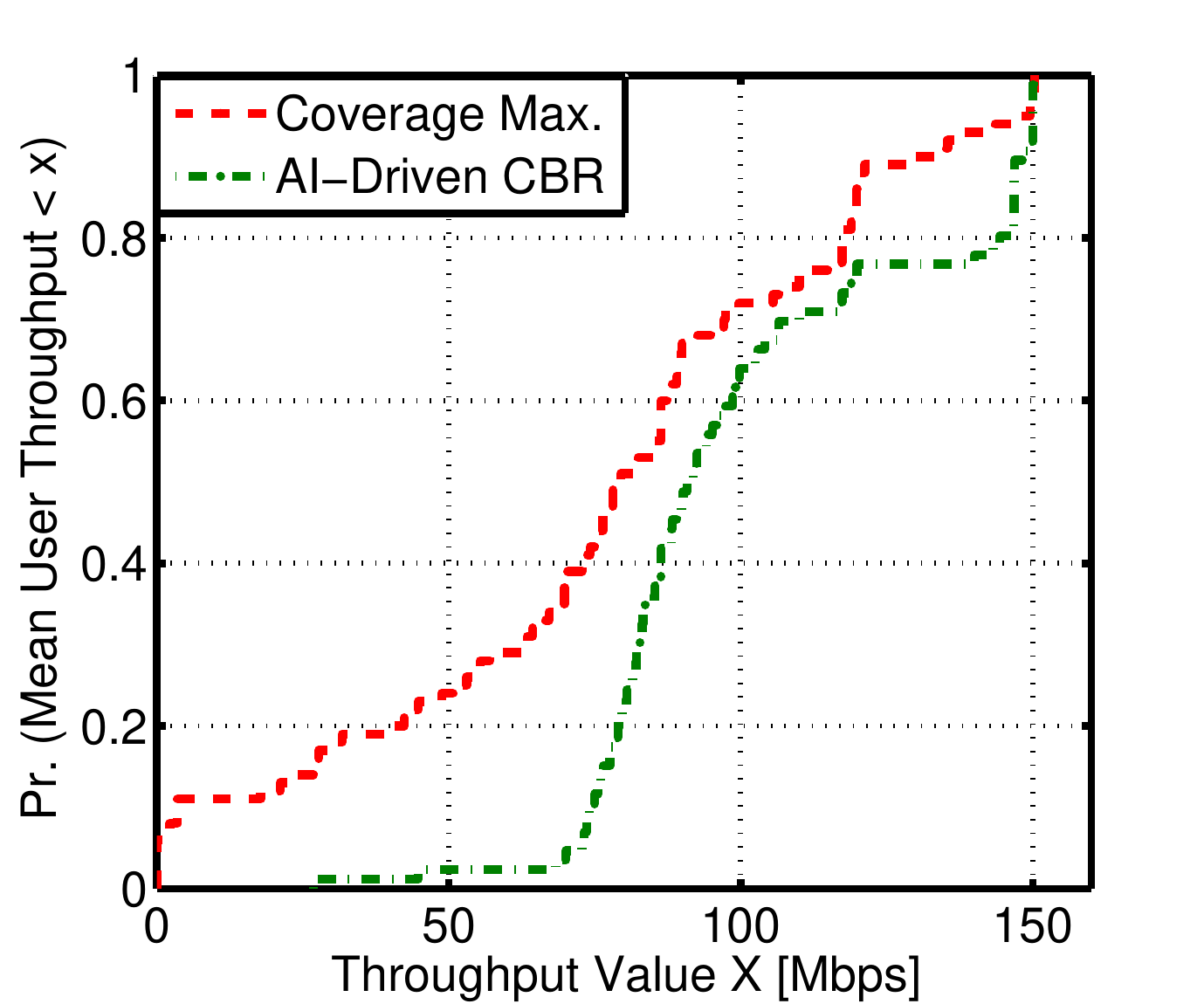}
	}
	\caption{Results of coverage problem in one apartment}
	\label{fig:SimRes1}
\end{figure*}

\textbf{\textit{Isolated Scenario:}} This scenario corresponds to simulating the coverage problem in an interference free environment, where the main source of throughput degradation is the poor RSSI. In \fref{fig:SimRes1}(a), the distribution of the throughput values by the three approaches is shown for the users connected to the extender which located far from the AP. While both the coverage maximization and AI self-deployment approaches have removed most of the service outages compared to the AP only scenario, the \emph{AI} obtained an average throughput of $ 88.5 $ Mbps with fairness index $ 0.95 $, while the coverage maximization resulted in {average throughput of} $ 77 $ Mbps and fairness index of $ 0.87 $. Nevertheless, the minimum throughput obtained by AI approach is $ 60 $ Mbps. At a higher user demand (i.e. $ 150 $ Mbps) in \fref{fig:SimRes1}(b), the AI gains increased over the traditional coverage maximization and resulted in no service outages, {average throughput of} $ 98.4$ Mbps, and {fairness index of }$ 0.91 $. On the contrary the coverage maximization resulted in $ 10 \%$ service outages, with $ 79 $ Mbps average throughput and $ 0.76 $ fairness index.

The above improvements demonstrate that the AI self-deployment is capable of 1) learning the throughput values without the need of modeling the walls, 2) reach the position that balances both back-haul and front-haul throughput values, and 3) remove the service outages and achieve uniform QoS across the users. This is unlike the coverage maximization approach that typically suffered from poor back-haul throughput and greedy placement that maximizes the front-haul throughput of some users over other farther ones.

\begin{figure*}[!t]
	\centering
	\subfigure[2 Apartments (2 STAs Each)]
	{
		\includegraphics[width=0.48\textwidth]{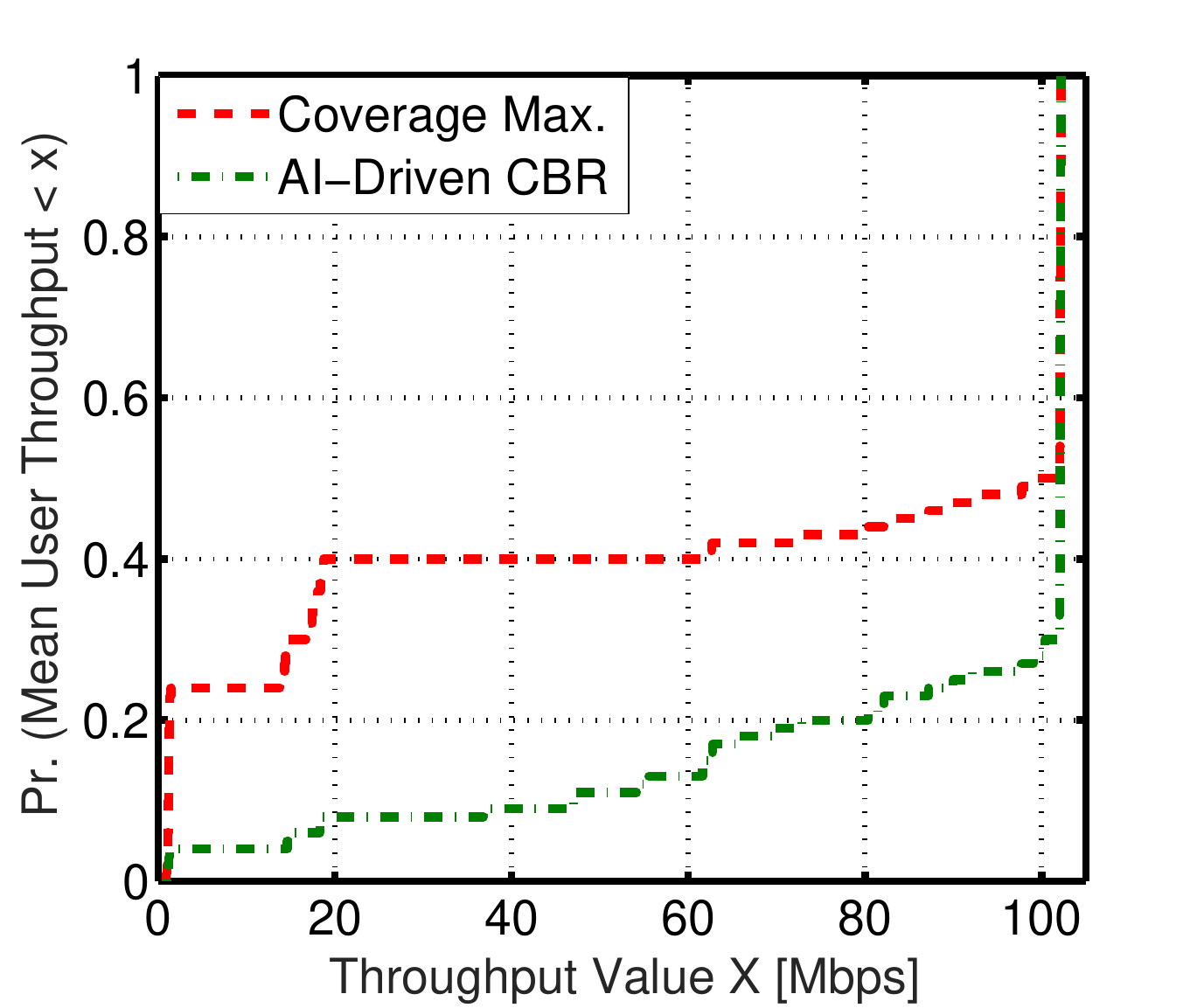}
	}
	\subfigure[1 Floor, 10 Apartments (1 STA Each)]
	{
		\includegraphics[width=0.48\textwidth]{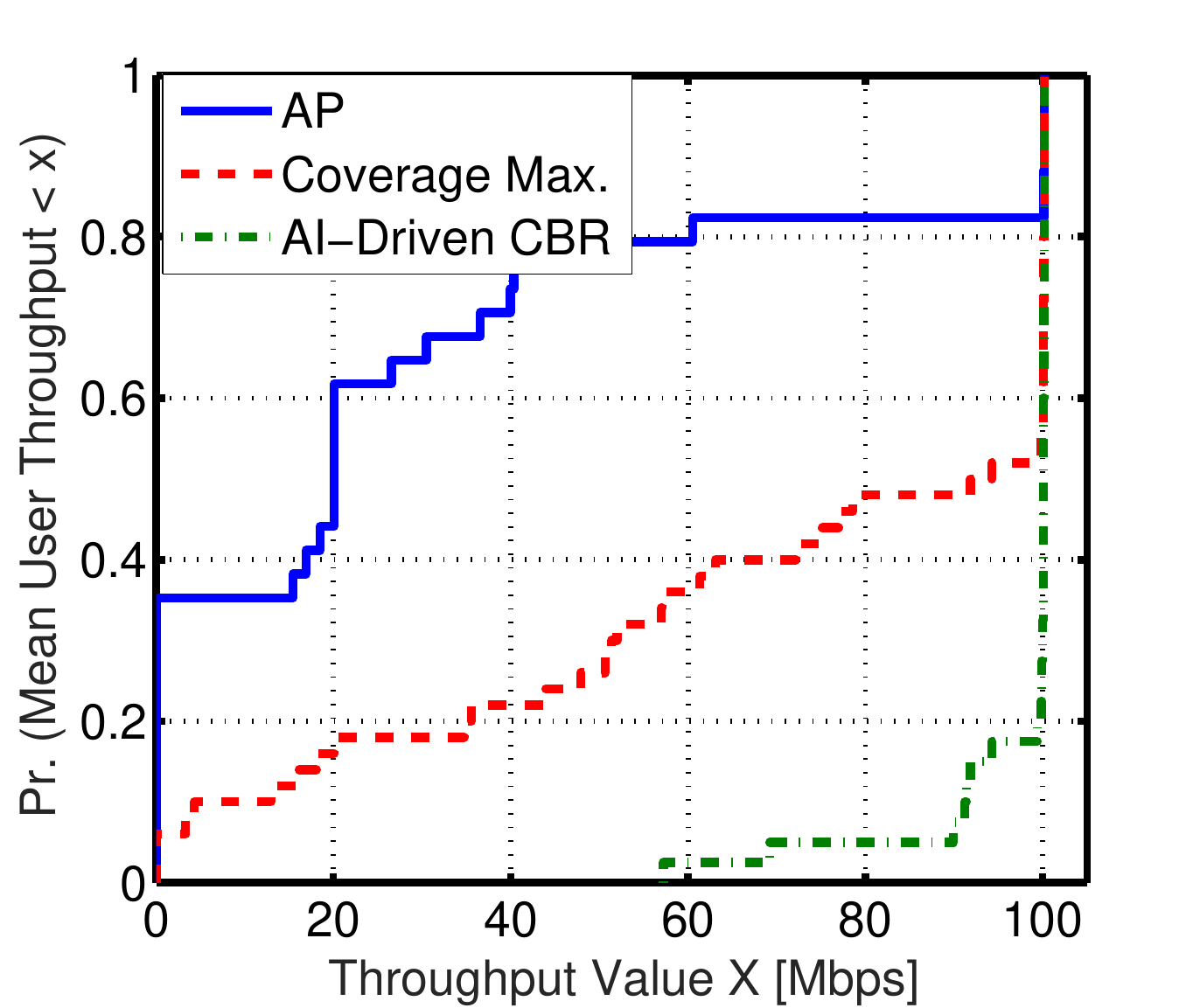}
	}
	\caption{Results of unmanaged apartment environment}
	\label{fig:SimRes2}
\end{figure*}

\textbf{\textit{Uncoordinated Scenario:}}
The simulations are further extended to uncoordinated scenario with one unmanaged neighbor as depicted in \fref{fig:SimScenarios}(a). The neighboring is adopting the same back-haul and front-haul channels as the managed apartment to simulate the worst case scenario. The two APs are placed such that they create a hidden node problem that increases the interference at the extender's back-haul. Similarly, the neighboring extender might create a hidden node problem at the front-haul when placed far away from the managed extender. The coverage maximization approach resulted in $ 25 \% $ service outages as depicted in \fref{fig:SimRes2}(a), in addition to an average throughput and fairness index of  $ 62.2 $ Mbps and $ 0.65 $, respectively.

The uncoordinated scenario is further extended to a single floor that comprises of one managed and nine unmanaged apartments as depicted in \fref{fig:SimScenarios} (b). The channels are randomly selected for all extenders and APs, while ensuring that different channels are assigned to the front-haul and back-haul of the same extender. The resultant throughput distribution of the managed users is reported in \fref{fig:SimRes2} (b). In this scenario, simultaneous hidden node problem on both back-haul and front-haul links is less frequent compared to the previous two apartments case due to the random channel assignment. As such, more	throughput improvements can be achieved by the AI driven approach that can minimize the impact of hidden node on back-haul without decreasing the signal level on the fronthaul link, and thus avoids hidden node in the latter when the extender is moved closer to the AP. The AI resulted in a minimum throughput of 60 Mbps and fairness index of 0.92. On the contrary, coverage maximization approach obtained less throughput values than 60 Mbps (minimum value by AI) in $ 40\% $ of the simulated cases.

\subsection{{Convergence}}
\begin{figure}[!t]
	\centering
	\includegraphics[width=0.9\linewidth]{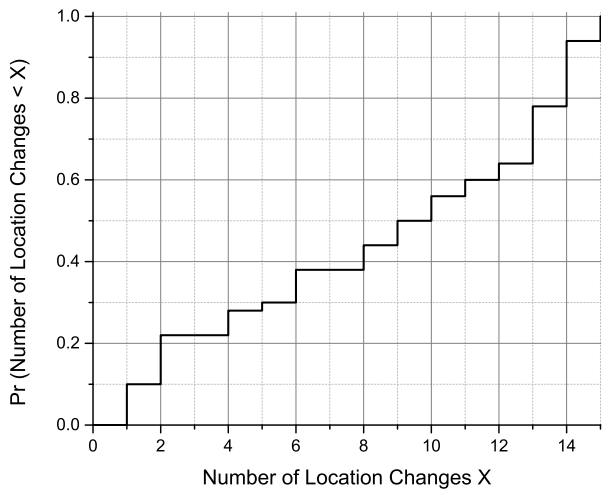}
	\caption{CDF of convergence test.}
	\label{fig:convergence}
\end{figure}

{For the scenario depicted in Fig. \ref{fig:SimScenarios}(a), we run 50 tests, with the same network parameters described above, where in each test we place the EXT and STA in random locations in the apartment and observe the number of location changes to converge to steady-state throughput. Then we plot the CDF of this distribution as given in Fig. \ref{fig:convergence}.}

{The mean of this distribution is 8.7 and the standard deviation is 4.9. From this information we can see that our framework converges at a sufficiently fast rate to be practical even in real-life networks.}

\section{Conclusions}
The ongoing shift in Wi-Fi indoor architecture from single-hop to multi-hop prompts more effort to achieve ubiquitous QoS. Optimal {self-deployment} of wireless extenders is introduced in this paper by adopting \ac{AI} that learns the network states and evaluates previous actions to reason new locations when the demanded \ac{QoS} is violated. The proposed AI framework optimizes the extenders' locations while implicitly taking into account the impact of uncoordinated neighboring networks and indoor obstacles on the user throughput. This is in addition to considering the trade-off between the back-haul (AP-to-extender) and front-haul (extender-to-user) throughput while evaluating each candidate location. The AI-framework is implemented on real testbed that demonstrates the feasibility of applying such advanced deployment strategy in practice. Using the testbed and standard compliant simulator ns-3, the {AI self-deployment} framework is evaluated in different indoor scenarios: both residential and enterprise with dense deployment of neighboring APs that create contention and interference. Compared to the state-of-the-art solutions, the proposed AI based approach achieves QoS fairness among the users, maximizes the average throughput value, and removes the service outage at distant users even in challenging uncoordinated scenarios where hidden node problem is substantial. These results support the momentum of applying {AI self-deployment} in future networks instead of the coverage maximization approaches proposed in the literature and used in today's networks. In the following we discuss practical aspects of extender self-deployment that have to be considered by system designers and operators while implementing AI self-deployment functionality.

\section{Future Directions}
These reported results on {AI-driven} self-deployment promise significant QoS improvements with minimal deployment cost. Here, we provide some future research directions that need to be addressed before implementation in practice. First, the user responsiveness to the network recommendations have to be modelled to 1) determine when the user is willing to re-position the extender and 2) compare the corresponding user-based location to the one requested by network. This modelling requires adopting machine learning in the perception and optimization stages, and results user-centric recommendations. A second aspect is improving the propagation model used in calculating the distance-based throughput (in the perception phase) to achieve faster convergence by the learning step. The third aspect is related to re-configuring the extender's and mAP's parameters such as operating channels at the new location. As such, joint optimization strategies are of paramount importance in order to avoid recommending locations with challenging radio conditions. The fourth aspect is modelling the error in localizing mAP, users and extenders, and incorporate the degree of uncertainty in the optimization step. {Another interesting future study may be evaluation of different learning techniques in this context}. Finally, cooperative learning among multiple managed scenarios (e.g. apartments) can be applied to transfer and reuse the existing knowledge base. Sensitivity analysis and robust techniques are also needed to handle uncertainties in the user responses and perception output. Nevertheless, guided heuristic techniques shall be proposed to guarantee scalable real-time solutions for the optimization problem at hand.

{The basic idea of the proposed framework is validated for a 2-hop network with both single and multiple stations by considering only the 2.4 GHz band. In most modern households, a single extender coupled with mAP will cover the performance needs of an overwhelming majority of users. As such, the demonstrated performance of the framework will serve as an important marker of the Quality of Service that can immediately be offered to these users through home wireless networks. However, as a future work, we are planning to thoroughly analyse the performance of framework in more complex scenarios such as in WMNs with multiple extenders with dual band radio (2.4 GHz and 5 GHz). Due to the markedly different nature of these bands, complexity of multi agent learning and multi objective optimization, it will be very challenging to treat the joint channel and location optimization problem in such type of WMNs.}

{We believe other multi-objective optimization formulation and multi-variable utility functions can be studied as well, yet have to be addressed separately as future work. Furthermore, due to existence of a wide variety of exploration strategies the study on their impact on optimizer performance is out of the scope of this work and it can be considered as interesting future study. Finally, due to many learning methods their study in this context may be another interesting work to pursue.}

\section*{Acknowledgement}
The authors would like to thank Mr. Samurdhi Karunaratne for his valuable comments and advices. We also thank to anonymous reviewers for their suggestions and comments that helped to improve the manuscript.

\section*{{Appendix A}}\label{appendix}
{Proof of Lemma 1 is given below.}
\begin{proof}
{First, we introduce the term of the capacitated facility location problem (CFLP), a well-known combinatorial optimization NP-hard problem} \cite{nphardCFLC}.

{In CFLP, we are given a set of clients and set of facilities. Each client has a demand which must be served by one or more open facilities. Fixed costs of opening facilities and the cost of serving a client demand by a certain facility are defined. The objective is to minimize the sum of fixed costs and transportation costs. Thus, our problem is equivalent with the definition CFLP where wireless extenders are equivalent to facilities and users are equivalent  to the clients in CFLP definition:}
\begin{itemize}
\item[1)] {the cost of deploying of the extender at location $i$ can be treated as a cost of opening the new facility;}
\item[2)] {the cost of repositioning of extender can be treated as the cost of transportation;}
\end{itemize}
{Since the formulation of our dynamic location optimization problem is equivalent to CFLP, we deduce that the defined extender relocation problem is NP-hard.}
\end{proof}


\end{document}